\documentclass[11pt]{article}

\usepackage{amssymb}
\usepackage{mathrsfs}
\usepackage{tikz}
\usepackage{chet}

\usetikzlibrary{calc,intersections,arrows.meta}

\tikzset{cross/.style={path picture={
      \draw[black]
            (path picture bounding box.south east) --
            (path picture bounding box.north west)
            (path picture bounding box.south west) --
            (path picture bounding box.north east);}}}

\newcommand{\cO}{\mathcal{O}}
\newcommand{\cH}{\mathcal{H}}
\newcommand{\dVM}{\Delta^{1/2}_{\text{VM}}}
\newcommand{\DeltaW}{\Delta^{\hspace{-2pt}\text{W}}}
\newcommand{\dsix}{\ensuremath{d^{\hspace{0.5pt}6}\hspace{-0.5pt}}}

\newcommand{\dtwo}{\ensuremath{d^{\hspace{0.5pt}2}\hspace{-0.5pt}}}
\newcommand{\dd}{\ensuremath{d^{\hspace{1pt}d}}}
\newcommand{\lsp}{\hspace{1pt}}
\newcommand{\llsp}{\hspace{0.5pt}}

\allowdisplaybreaks

\date{April 2015}

\preprint{UCSD/PTH 05-01}

\title{Two\hspace{1.5pt}-loop renormalization of multiflavor
$\titlemath{\phi^3}$ theory\\\vspace{4pt}
in six dimensions and the trace anomaly}

\author{Benjam\'in Grinstein,$\!^{a}$
Andreas Stergiou,$\!^{b}$
David Stone$^{c}$ and
Ming Zhong$^{d}$
\emails{\href{mailto:bgrinstein@uscd.edu}{bgrinstein@uscd.edu},
\href{mailto:andreas.stergiou@yale.edu}{andreas.stergiou@yale.edu},
\href{mailto:david.curtis.stone@roma1.infn.it}{david.curtis.stone@roma1.infn.it},
\href{mailto:zhongm@nudt.edu.cn}{zhongm@nudt.edu.cn}}}

\affiliation{$^a$Department of Physics, University of
California, San Diego, La Jolla, CA 92093, USA\\
$^b$Department of Physics, Yale University, New Haven,
CT 06520, USA\\
$^c$INFN, Sezione di Roma, Piazzale A.\ Moro 2, I-00185 Roma, Italy\\
$^d$Department of Physics, National University of Defense
Technology, Hunan 410073, China}

\abstract{We use the background-field method and the heat kernel to obtain
all counterterms to two-loop order of conformally-coupled multiflavor
$\phi^3$ theory in six spacetime dimensions, defined in curved spacetime and
with spacetime-dependent couplings. We also include spacetime-dependent
mass terms for completeness. We use these results to write a general
expression for the trace anomaly. With the use of Weyl consistency
conditions we are able to show that the strong $a$-theorem for a certain
natural candidate quantity $\tilde{a}$ is violated in this theory, and
obtain a three-loop expression for the coefficient $a$ of the Euler term in
the anomaly.}

\begin{document}

\maketitle

\toc

\newsec{Introduction}
Classical field theories that are invariant under scale and special
conformal transformations generally fail to retain these symmetries once
quantized. Famously there is an anomaly, that is, the trace of the
stress-energy tensor does not vanish, signaling a violation of invariance
by rescalings. The exception consists of a class of quantum field theories
for which the trace vanishes, known as conformal field theories (CFTs).
When this happens not only is scale invariance restored, but the theory is
also symmetric under the full group of conformal
transformations~\cite{Fortin:2012hn, Luty:2012ww, Dymarsky:2013pqa,
Dymarsky:2014zja}. This occurs at the fixed points of the renormalization
group (RG) flow.

It is also of interest to put the quantum field theory of interest on a
curved background. When quantizing such a theory there are trace anomalies
even at the fixed points of the RG flow of the corresponding flat-space
theory~\cite{Capper:1974ic, Duff:1993wm}.  These anomalies are given by a
diffeomorphism-invariant local function involving derivatives of the
metric. In $d$ dimensions, there are a finite number of contributions of
mass dimension $d$; for each there is a coefficient which is a function of
the couplings. These coefficients are often of interest. Most notably, the
coefficient $a$ of the $d$-dimensional Euler density is Cardy's proposed
extension~\cite{Cardy:1988cwa} of the central charge $c$ of two-dimensional
CFTs, whose monotonicity properties under RG flow were understood by
Zamolodchikov~\cite{Zamolodchikov:1986gt}.

These coefficients, some of which are the central charges of the theory,
are well-understood at the fixed points but are also defined along the RG
flow. In two dimensions, a
suitable extension of $c$ away from fixed points may be defined, called
$\tilde{c}$, which is a function of the couplings, so
one may speak of their values along the RG flow in a sensible fashion. It
is this quantity, $\tilde{c}$, that has the interesting properties that it
decreases monotonically along RG flows and is stationary at fixed points
where it takes the numerical value of the central charge $c$ of the CFT
corresponding to the fixed point.

Given such remarkable properties of $\tilde{c}$, it is natural to ask
whether such a quantity exists in the more physically interesting
four-dimensional case. In fact, Weyl consistency
conditions~\cite{Osborn:1991gm, Grinstein:2013cka} identify a quantity
$\tilde{a}$ in even spacetime dimensions that make it the one possible
candidate for a generalization of Zamolodchikov's $\tilde{c}$ to higher
dimensions. In four dimensions it was shown by Jack and
Osborn~\cite{Jack:1990eb} that this quantity is stationary at fixed points
where it reduces to the coefficient $a$ of the Euler term. Moreover, using
perturbation theory they showed that this quantity is monotonically
decreasing towards the IR. More specifically, they gave an equation for the
RG flow of $\tilde{a}$ that implies its monotonicity if a certain symmetric
tensor, or ``metric'' in theory space parametrized by the couplings of the
theory, is positive-definite.  They then showed in an explicit perturbative
calculation that this metric is in fact positive-definite for small
couplings. More recently, positivity of this metric has been established in
conformal perturbation theory \cite{Baume:2014rla, Osborn:2015rna}.

The extension of the quantity $\tilde{a}$ to six dimensions was computed by
a set of the current authors in~\cite{Grinstein:2013cka}, and furthermore was
shown to have a natural definition in any even-dimensional spacetime as a
consequence of the Weyl consistency conditions and the existence of a
generalization of the Einstein tensor, along with a metric on the space of
couplings that is analogous to that of Jack and Osborn and Zamolodchikov.
This generalization of $\tilde{a}$ is stationary at fixed points and
reduces to $a$ there. However, surprisingly, in~\cite{Grinstein:2014xba} we
showed by explicit computation in perturbation theory for a theory of
scalars with a cubic self-coupling that the metric is negative-definite,
and so $\tilde{a}$ monotonically increases in the flow out of the trivial
UV fixed point.  Adding to this surprise, in~\cite{Osborn:2015rna} it was
found that in a model with two-forms in six dimensions the metric is
positive-definite. It seems that, even in perturbation theory, there is no
straightforward generalization of the $a$-theorem in six dimensions, at
least as envisioned in the cases so far. As explained
in~\cite{Osborn:2015rna}, this may be attributed to the fact that in six
dimensions the trace anomaly on a conformal manifold defines three
independent symmetric tensors on the space of couplings, only one of which
satisfies positivity properties. This positive-definite tensor is, however,
not the tensor that appears in the RG equation for $\tilde{a}$, and thus
the monotonicity of its flow remains undetermined. Contrary to this, in two
and four dimensions there is a unique symmetric tensor with established
positivity properties that also appears in the RG equation for $\tilde{c}$
or $\tilde{a}$.

It is not known beyond perturbation theory whether flows of $\tilde{a}$ in
four dimensions are monotonic.  However, there is another approach to the
$a$-theorem that does not follow the previous lines of computation that
uses unitarity of scattering processes in dilaton effective theories to
establish positivity. Komargodski and Schwimmer have
argued~\cite{Komargodski:2011vj} without recourse to perturbation theory
that the value of $a$ on the UV fixed point is larger than that at the IR
fixed point.\foot{This is a statement of the weak version of the
$a$-theorem in four dimensions, where it is only established that
$a_{\text{UV}} > a_{\text{IR}}$. The strong version of the $a$-theorem
states that $\tilde{a}$ decreases monotonically along the flow. The
strongest version of the $a$-theorem states that this flow is a gradient
flow~\cite{Barnes:2004jj}.}  A similar argument considered the same
question in six dimensions~\cite{Elvang:2012st}; however it was not
possible to reach a conclusion with the same methods as Komargodski and
Schwimmer. Perhaps related in a general way to the difficulties encountered
in~\cite{Elvang:2012st}, we note that the (massless) scalar model with cubic
interactions investigated in~\cite{Grinstein:2014xba} has only a single
Gaussian (trivial) fixed point within the domain of validity of the
perturbative calculation, so the difference between the values of $a$ in
the UV and IR cannot be contemplated. Non-perturbative CFTs are known to
exist in six dimensions, but since in addition to being non-perturbative
they are non-Lagrangian CFTs, little is known about flows between them.

As is clear from our discussion so far, the situation in six dimensions is
significantly more complicated than that in two and four dimensions.  We
believe it would be useful to gain as much information as possible about
the perturbative behavior of six-dimensional theories, beyond the
computation of the quantity $\tilde{a}$. With this motivation in mind, we
compute in this work, at two loops, the infinite part of the effective
action and the trace anomaly in multiflavor $\phi^3$ theory in six
dimensions, including, for completeness of the analysis, the possibility
that scale invariance is explicitly broken classically by a mass term.  In
addition to computing in a curved background (with a spacetime-dependent
metric) we take the couplings to also have spacetime dependence. In effect
this allows us to study the renormalization of correlators of operators
that appear in the Lagrangian. With spacetime-dependent couplings
counterterms
proportional to derivatives of the couplings are required for finiteness.
Correspondingly, the trace anomaly includes terms that contain derivatives
not just of the metric but also of the couplings.  The anomalies associated
with these terms manifest themselves in the original model (with
spacetime-independent metric and couplings) as coefficients of terms in the
Greens' functions of composite operators (including the stress-energy
tensor and its trace).

Given all these considerations, the focus of this paper is the Lagrangian
\eqn{\mathscr{L}=\tfrac12\big(\partial_\mu\phi_i\,\partial_\nu\phi_i
\,\gamma^{\mu\nu} + (\xi_{ij} R+m_{ij}) \lsp\phi_i \phi_j\big)+h_i\phi_i
+ \tfrac{1}{3!} g_{ijk} \phi_i\phi_j\phi_k\,,}[lagrangian]
of scalar fields $\phi^i$, defined on a six-dimensional manifold with
metric $\gamma^{\mu\nu}$, where the repeated lowercase Latin flavor indices
are to be summed over regardless of their position. This Lagrangian is of
interest because it is the only general interacting theory that has
classical scale invariance (for the appropriate choice of $\xi_{ij}$ and
zero $m_{ij}$ and $h_i$) in six dimensions.\foot{One may object that the
$\phi^3$ theory is sick because of its potential, which is unbounded from
below. However, within the context of perturbation theory, which is the
scope of this paper, the ground state $\langle \phi (x) \rangle = 0$ is
stable to fluctuations of $\phi(x)$~\cite{Rubakov:2002fi}.} Of course, one
may consider theories that do not have Lagrangian descriptions in six
dimensions, as mentioned above, but then the calculational methods of this
paper are of no use and, typically, one must resort to holographic methods.
With \lagrangian we may proceed in the old-fashioned ways of perturbation
theory and reliably calculate the quantities of interest order by order in
$g_{ijk}$. This is the starting point of this paper but first we must
establish how such computations are performed on curved backgrounds. We
should note that our results are reported here with the choice
$\xi_{ij}=\frac15\delta_{ij}$ classically, with $\delta_{ij}$ the Kronecker
delta, as found from the general result
$\xi_{ij}=\frac{d-2}{4(d-1)}\delta_{ij}$ in $d$ dimensions for conformal
coupling of the scalar.

The main computational method used in this work was developed and applied
to various cases in four dimensions by Jack and Osborn in
\cite{Jack:1982hf, Jack:1982sr, Jack:1983sk, Jack:1984vj}. The main
ingredients are the background field method and the heat kernel in
dimensional regularization. In $\phi^3$ theory in six dimensions with a
single scalar field and with spacetime-independent couplings,  results for the
two-loop effective action have been
obtained in~\cite{Jack:1985wf, Kodaira:1985vr, Kodaira:1985pg}; we have
checked our results for some quantities against those listed in these references.

The layout of this paper is as follows. In the next section we describe in
detail the (perhaps unfamiliar but very powerful) computational method of
Jack and Osborn. In section \ref{weylCCsSec} we describe briefly the Weyl
consistency conditions in order to make contact between our computations of
the effective action and the $a$-theorem. In section \ref{polesEffAction}
we present our results for the infinite part of the effective action at two
loops, and in section \ref{betagammasec} we extract from those the two loop
beta function and anomalous dimension. Finally, in section \ref{metricSec}
we present results relevant to the $a$-theorem in six dimensions to
three-loop order. Our conventions as well as various details and results
needed for our computations are contained in three appendices.

\textbf{Note added in v2:} The discussion in sections 3 and 6 contains
errors that have been addressed in~\cite{Stergiou:2016uqq}.

\newsec{Method of calculation}[methodOfCalc]
In this section we outline the method of calculation employed in this
paper. For more details the reader is referred to~\cite{Jack:1982hf,
Jack:1982sr, Jack:1983sk, Jack:1984vj}, where such computations have been
thoroughly explained and demonstrated. Until section \ref{polesEffAction}
we assume for simplicity that no relevant parameters are present, for
example $m_{ij} = 0$ and $h_i = 0$ in \lagrangian.

In this work we will study quantum field theories defined in spacetime
dimension $d=D-\epsilon$, with $D$ an integer, by a set of couplings
$g^{\prime\lsp I}$ and fields $\phi^{\prime}$. For our computations we
will use dimensional regularization and make explicit the mass dimension of
the renormalized parameters via
\eqn{g^{\lsp\prime\lsp I}=\mu^{k^I\epsilon}g^{I}\,,\qquad
\phi^{\lsp\prime}=\mu^{\delta\epsilon}\phi\,,}[gphiMassD]
for some numbers $k^I$ and $\delta$ and where the index $I$ labels the
operators in the interaction Lagrangian, i.e.\ $I = (ijk)$ in
\lagrangian.\foot{Note that the index carried by $k$ of \gphiMassD is not
subject to the summation convention.} Though we start the perturbative
calculations with $g^{\lsp\prime\lsp I}$ and $\phi^{\lsp\prime}$, we will
use \gphiMassD to express the resulting formulas in terms of the fields
$\phi$ and the dimensionless couplings $g^{I}$. Then, with minimal
subtraction, we have the bare parameters
\eqn{g_0^I=\mu^{k^I\epsilon}(g^I+L^I(g))\,,\qquad
\phi_0=\mu^{\delta\epsilon}Z^{1/2}(g)\phi\,,}[BareToRen]
with $L^I$ and $Z^{1/2}-1$ containing just poles in $\epsilon$.  In general
$Z^{1/2}$ is a matrix to account for the multiple number of fields in
\lagrangian. The beta
function and anomalous dimension are given by
\eqn{\hat{\beta}^I\equiv\mu\frac{dg^I}{d\mu}=-k^Ig^I\epsilon+\beta^I\qquad\text{and}\qquad \hat{\gamma}\equiv \delta\epsilon
+Z^{-1/2}\mu\frac{dZ^{1/2}}{d\mu}=\delta\epsilon+\gamma\,,}[]
respectively, where $\beta$ and $\gamma$ are the quantum beta function and
anomalous dimension respectively.

Now, in quantum field theory in flat spacetime, wavefunction and coupling
renormalization are enough to render finite correlation functions involving
fundamental fields. When correlation functions involving composite
operators are included, further counterterms are necessary. A convenient
way to deal with these is by introducing sources for the composite
operators, and including counterterms proportional to spacetime
derivatives on those sources. For operators that appear in the Lagrangian
it is enough to take their couplings as spacetime-dependent sources,
$g^I\to g^I(x)$, and introduce counterterms proportional to derivatives on
$g^I(x)$~\cite{Jack:1990eb, Shore:1990wp}. Finally, when a flat-space field
theory is lifted to curved space with metric $\gamma_{\mu\nu}$, and the
regularization procedure respects diffeomorphism invariance, new
divergences proportional to the curvatures defined from $\gamma_{\mu\nu}$
appear, and thus further counterterms involving the curvatures are required
for finiteness.

In \cite{Jack:1990eb} a systematic treatment of such effects was
undertaken, and a general expression for the Lagrangian in the presence
of the sources $\gamma_{\mu\nu}(x)$ and $g^I(x)$ was proposed, namely
\eqn{\tilde{\mathscr{L}}_0 = \mathscr{L}_0
-\mu^{-\epsilon}\lambda\cdot\mathscr{R}+\mu^{-\epsilon}
\mathscr{F}\,,}[Ltilde]
where $\lambda\cdot\mathscr{R}$ includes all field-independent
counterterms, proportional only to curvatures and derivatives on $g^I(x)$,
and $\mathscr{F} = \mathscr{F}(\phi)$ includes all field-dependent
counterterms that also depend on curvatures and derivatives on $g^I(x)$.
$\mathscr{L}_0$ is the bare Lagrangian, expressed in terms of $g$ and
$\phi$ with the use of \BareToRen, that contains terms that survive in flat
space when the couplings are taken to be spacetime independent. It obeys
the Callan--Symanzik equation
\eqn{\left( \hat{\beta}^I \frac{\partial}{\partial g^I} -
(\hat{\gamma}\phi)\cdot \frac{\partial}{\partial \phi} - \epsilon \right)
\mathscr{L}_0 = 0\,.}[CZeqn]

The RGE one finds from $\Ltilde$ is
\eqn{\left(\hat{\beta}^I\frac{\partial}{\partial g^I}
-(\hat{\gamma}\phi)\cdot\frac{\partial}{\partial\phi}-\epsilon\right)
\tilde{\mathscr{L}}_0=\mu^{-\epsilon}\left(\beta_\lambda\cdot\mathscr{R}+
\left(\hat{\beta}^I\frac{\partial}{\partial g^I}
-(\hat{\gamma}\phi)\cdot\frac{\partial}{\partial\phi}-\epsilon\right)
\mathscr{F}\right),}[RGEFull]
which, by \Ltilde and the Callan--Symanzik equation \CZeqn, requires
\eqn{\left(\epsilon-\hat{\beta}^I\frac{\partial}{\partial g^I}\right)
\lambda\cdot\mathscr{R}=\beta_\lambda\cdot\mathscr{R}\,,}[RGEct]
and similarly for the $\mathscr{F}(\phi)$ terms, though there is an
additional derivative with respect to the fields. As explained
in~\cite{Jack:1990eb} and we will review in the following, the terms
$\beta_\lambda\cdot \mathscr{R}$ defined by \RGEct contribute, among
others, to the trace anomaly of the theory in curved space.

It is important to emphasize that in specific theories with possible
relevant parameters like \lagrangian, the RGE \RGEFull is incomplete. For
example, it does not correctly reproduce higher-order poles in higher-loop
computations, even if the relevant parameters are set to zero in the
classical Lagrangian. This issue has been analyzed in detail in
\cite{Jack:1990eb} for four-dimensional theories, and also in
\cite{Osborn:2015rna} for \lagrangian. While it does not affect our
discussion below, it should be kept in mind.

\subsec{Background field method}[bfmethod]
In this subsection we will give a brief overview of the background field
method. We will present our expressions for the case of a single scalar
field $\phi$, although the generalization to multiple fields and fields
with spin is well-known. Our motivation for using the background field
method is that it allows us to compute perturbatively counterterms like
$\lambda\cdot\mathscr{R}$ in \Ltilde in a straightforward way.

In the background field method one simply computes the effective action
starting from $\mathscr{L}_0$, which thus dictates the form of the
counterterms. More specifically, we start by splitting the field $\phi$
into an arbitrary classical background part $\phi_b$ and a quantum
fluctuation $f$,
\eqn{\phi=\phi_b+f\,.}[]
We can also introduce a source $J$, and obtain the effective action
$W[\phi_b,J]$ (the generating functional of connected graphs with
implicit $\gamma_{\mu\nu} (x)$ and $g^I(x)$ dependence) after we
integrate out $f$:
\eqn{e^{W[\phi_b,J]}=\int D\hspace{-0.8pt}f\,
e^{-\tilde{S}_0[\phi]+\int\dd x\sqrt{\gamma}\,J(x)f(x)}\,,\qquad
\tilde{S}_0=\int\dd x\sqrt{\gamma}\,\tilde{\mathscr{L}_0}\,,}[EffAction]
where $\gamma$ is the determinant of the metric $\gamma_{\mu\nu}$, which is
not to be confused with the anomalous dimension $\gamma$.

To continue, let us denote by $S^{(0)}$ the action without any
counterterms. Then, we expand $S^{(0)}[\phi]$ in fluctuations,
\eqn{S^{(0)}[\phi]=S^{(0)}[\phi_b]+\int\dd
x\sqrt{\gamma}\left.\frac{\delta S^{(0)}}{\delta\phi}
\right|_{\phi_b}f+\frac12\int \dd x\sqrt{\gamma}\,fMf+S_{\text{int}}[f]\,,
}[fluc]
where $M=-\nabla^2 + \left. \dtwo V/d\phi^2 \right|_{\phi = \phi_b}$, with $V$ the potential in
$\mathscr{L}$.  Then, by expanding \EffAction we find that, at the zeroth
order,
\eqn{W^{(0)}[\phi_b]=-S^{(0)}[\phi_b]\,,}[]
and at the one-loop order (a superscript in parentheses indicates the loop
order),
\eqn{W^{(1)}[\phi_b]=-\tilde{S}_0^{(1)}[\phi_b]-\tfrac12
\ln\det M\,,}[OneLoopEffAction]
after we choose $J$ appropriately in order to cancel terms linear in $f$,
order by order in perturbation theory starting with \fluc, and subsequently
perform in \EffAction the Gaussian integral over $f$. Here,
$\tilde{S}_0^{(1)}$ contains poles in $\epsilon$ to cancel those in the
$-\tfrac12\ln\det M$ piece; in particular it contains the one-loop
contributions to $Z^{1/2}$ and $L$ of \BareToRen, which are chosen to
absorb the associated infinities coming from $-\tfrac12\ln\det M$ so that
$W^{(1)}$ is finite. In addition, with the extension \Ltilde it is clear
from $\OneLoopEffAction$ that $\tilde{S}_0^{(1)}$ also contains the
one-loop contribution to $\lambda\cdot\mathscr{R}$ that is given by the
negative of the appropriate simple-pole part of $-\tfrac12\ln\det M$:
\eqn{\int\dd x\sqrt{\gamma}\,\mu^{-\epsilon}\lambda^{(1)}\cdot{R}\subset
-(-\tfrac12\ln\det M)^{\text{pole}}\,.}[OneLoopCTlR]
Then, from \RGEct and \OneLoopCTlR we can evaluate
$\beta_\lambda^{(1)}\cdot\mathscr{R}$. Of course, $-\tfrac12\ln\det M$ also
contains field-dependent terms that require the counterterms
$\mathscr{F}^{(1)}$ for finiteness.

At higher loops the interaction term $S_{\text{int}}[f]$ in \fluc is
considered and vacuum bubble diagrams as well as diagrams with counterterm
insertions are constructed.\foot{As explained in~\cite{Jackiw:1974cv}, one
of the advantages of computations done in the background field method is
that only vacuum bubble diagrams need be considered order by order in
perturbation theory.} The counterterms are of course fixed here by the
previous loop order, i.e.\ by $\tilde{S}_0^{(1)}$.  These diagrams can be
evaluated in position space, using coincident limits of propagators
according to the diagram topology. With these methods, which are explained
thoroughly in the following, no loop integrations need to be performed. If
we denote by $\mathscr{S}^{(2)}$ the contribution of all such diagrams, we
find
\eqn{W^{(2)}[\phi_b]=-\tilde{S}_0^{(2)}[\phi_b]
+\mathscr{S}^{(2)}\,.}[]
Again, finiteness of $W^{(2)}$ allows us to determine all counterterms in
$\tilde{S}_0^{(2)}$.  From the simple poles in
$\lambda^{(2)}\cdot\mathscr{R}$ it is again straightforward to evaluate
$\beta_\lambda^{(2)}\cdot\mathscr{R}$ using the RGE \RGEct. Clearly these
computations can be carried out order by order in perturbation theory.

\subsec{Heat kernel}[secHeatKernel]
Using heat-kernel techniques the evaluation of $(-\tfrac12\ln\det
M)^{\text{pole}}$ and higher loop poles may be accomplished. A pedagogical
explanation of the method may be found in~\cite{Mukhanov:2007zz}; we will
mainly follow the procedure as laid out in~\cite{Jack:1985wf}, where it was
used for single flavor $\phi^3$ theory without spacetime-dependent
couplings. A nice review of the heat kernel method and its applications can
be found in \cite{Vassilevich:2003xt}.

The object of central importance in the heat kernel method is the
propagator function in the presence of a background field as presented in
section~\bfmethod. It obeys the identity
\eqn{ M_{ik}\lsp G_{kj} (x,x') = \delta_{ij}\lsp\delta^{\llsp d}
(x,x')\,,}[greensfunction]
where the indices are the flavor indices of the theory \lagrangian,
$\delta_{ij}$ is the Kronecker delta, the $d$-dimensional biscalar delta
function is defined by
\eqn{\int \dd x'\sqrt{\gamma'}\,\delta^{\llsp
d}(x,x')\phi(x')=\phi(x)\,,\qquad \gamma'=\gamma(x')\,,}[]
and $M_{ij}$ is the elliptic differential operator, evaluated at the point
$x$, alluded to in the previous section and defined by \lagrangian in our
case of interest, having the general form
\eqn{M_{ij} = -\delta_{ij} \nabla^2 + \left. \frac{\partial^2
V(\phi)}{\partial \phi^i\,\partial \phi^j}\right|_{\phi=\phi_b}\,.}[]
The key to evaluating the determinant in the one-loop effective potential
\OneLoopEffAction and the higher order diagrams, which involve integrals
over products of $G_{ij}(x,x')$, is to present $G_{ij}(x,x')$ in a
way amenable to computation.

The heat kernel provides such amenities. First, we define the heat kernel
$\mathscr{G}_{ij}$ by the equation
\eqn{ \left( \delta_{ik} \frac{\partial}{\partial t} + M_{ik} \right)
\mathscr{G}_{kj} (x,x';t) = 0\,, }[heateqn]
with
\eqn{\mathscr{G}_{ij}(x,x';0)=\delta_{ij}\lsp\delta^{\llsp d}(x,x')\,. }[initcondition]
Formally, by virtue of \heateqn, the heat kernel may then be written
\eqn{\mathscr{\hat{G}} (t)
= e^{-\hat{M} t} =
\sum_n e^{-\lambda_n t} | \psi_n\rangle \langle \psi_n |\,,
}[formalheateqn]
with $\lambda_n$ the eigenvalues of $\hat{M}_{ij}$ and the hats emphasizing
that no particular basis of eigenstates $| \psi_n \rangle $ for the
elliptic differential operator $\hat{M}_{ij}$ need be chosen (however, the
position basis will recover our calculations).
$G_{ij} (x,x')$ may be then written as
\eqn{G_{ij} (x,x') =
\int_0^{\infty}dt\,\mathscr{G}_{ij}(x,x';t)\,.}[greensFxnHeatKernel]

From the heat kernel $\mathscr{G}_{ij} (x,x';t)$ the one-loop effective
action is obtained through the well-established zeta-function method
(elaborated in, e.g.,~\cite{Jack:1983sk} or~\cite{Mukhanov:2007zz}), which
relates $-\tfrac{1}{2} \ln \det M$ to the heat kernel; to do so, an Ansatz
for the form of the heat kernel must be given. This is suggested from the
solution in flat space for the heat equation \heateqn and was given by
DeWitt~\cite{DeWitt:1965jb} for a small $t$ expansion:
\eqn{ \mathscr{G}_{ij} (x,x'; t) = \frac{\Delta_{\text{VM}}^{1/2}(x,x')}
{(4\pi t)^{d/2}}e^{\sigma (x,x')/2t} \sum_{n=0}^\infty
a_{n,\lsp ij} (x,x') \lsp t^n\,,\qquad
a_{0,\lsp ij}(x,x)=\delta_{ij}\,,}[hkansatz]
with $a_{n,\lsp ij}(x,x')$ the so-called Seeley--DeWitt
coefficients and where $\sigma(x,x')$ is the biscalar distance-squared
measure (called the geodetic interval by DeWitt),
\eqn{\sigma(x,x') = \tfrac{1}{2} \left( \int_0^1 d\lambda \,
\sqrt{\gamma_{\mu\nu} \frac{dy^{\mu}}{d\lambda}\lsp
\frac{dy^{\nu}}{d\lambda}} \, \right)^2 \,,\qquad y(0)=x\,,\,\,y(1)=x'\,,}[]
with $y(\lambda)$ a geodesic. $\Delta_{\text{VM}}(x,x')$ is another
biscalar, called the van Vleck--Morette determinant,  that describes the
spreading of geodesics from a point, defined by
\eqn{\Delta_{\text{VM}} (x,x') =\gamma(x)^{-1/2}\lsp\gamma(x')^{-1/2}
\det\left(-\frac{\partial^2}{\partial x^{\alpha}
\partial x'^{\beta}} \sigma (x,x') \right). }[]
We shall suppress the $x$, $x'$ dependence of $\sigma$,
$\Delta_{\text{VM}}$, and $a_{n,\lsp ij}$ henceforth. Now, the Ansatz
\hkansatz obeys \heateqn which yields the recursion relation
\eqn{n\hspace{0.5pt} a_{n,\lsp ij}+\partial_{\mu}\sigma\,
\partial^{\mu}a_{n,\lsp ij} =
-\Delta_{\text{VM}}^{-1/2} M_{ik} \big(\Delta_{\text{VM}}^{1/2}
a_{n-1,kj}\big)  \qquad \text{with} \qquad
\partial_{\mu}\sigma\,\partial^{\mu}a_{0,\lsp ij}=0\,,}[recursionrelation]
which allows us to compute the Seeley--DeWitt coefficients.

With the asymptotic expansion of the propagator via the heat kernel
established in \hkansatz, its practicality in loop computations becomes
evident. To elaborate on the comments above \hkansatz, at one loop one
wishes to calculate the determinant in \OneLoopEffAction. This may
accomplished by considering the so-called zeta function for the operator
$M_{ij}$,
\eqn{ \zeta_{M} (s) = \frac{1}{\Gamma (s) } \int_0^{\infty} dt \,
t^{s-1} \int \dd x \sqrt{\gamma} \, \mathscr{G}_{ii} (x,x; t)\,. }[zetaPositionBasis]
This function is useful to define because then the log of the determinant
may be computed by differentiating it with respect to $s$ and sending $s$
to zero, which may be seen by considering the formal definition of
$\mathscr{G}_{ij} (t)$ in \formalheateqn; this yields
\eqn{-\ln \det M = \lim_{s \to 0} \frac{d \zeta_{M}}{ds} =
\int_0^{\infty} \frac{dt}{t}
\, \int \dd x \sqrt{\gamma} \,\mathscr{G}_{ii} (x,x; t)\,.
}[zetaDeterminantRelation]
Given the formal definition in \formalheateqn, the value of $\ln
\det M = \sum_n \ln \lambda_n$ may computed with the equivalent of
\zetaPositionBasis for $\hat{\mathscr{G}}$, with $\text{Tr} \, \hat{\mathscr{G}}
(t) = \sum_n e^{-\lambda_n t}$.  Explicitly evaluating $\zeta_M$ as a function
of $s$, differentiating with respect to $s$ and then taking the limit as $s
\rightarrow 0$ reproduces the log of the determinant, formally, up to a
minus sign.

Actually, the equality in \zetaDeterminantRelation is only true up to the
residue of a pole in $s$ as $s
\rightarrow 0$ and equation \zetaDeterminantRelation is a bit misleading at
face value.  Following~\cite{Jack:1983sk}, $\lim_{s \to 0} \zeta_M^{\lsp
\prime} (s)$ is of the form
\eqn{ \lim_{s \rightarrow 0} \frac{d \zeta_M}{ds} = \lim_{s \rightarrow 0}
\left( \int_0^{\infty} dt \, t^{s-1} \int \dd x
\sqrt{\gamma} \, \mathscr{G}_{ii} (x,x;t) \; - \;
\frac{P}{s} \right)\,, }[]
with $P$ the residue of the integral inside the parentheses as $s
\rightarrow 0$. However, in dimensional regularization this pole is
displaced---this may be seen by noting the $d$ dependence of the power
series in $t$ in \hkansatz. Hence, in dimensional regularization, $P$ may
be set to zero and we recover the $\epsilon$-dependent determinant
\eqn{\left( - \ln \det M \right)_{\text{dim.} \, \text{reg.}} =
\left. \lim_{s \rightarrow 0} \frac{d \zeta_M}{ds}\right|_{d = D- \epsilon}  =
\int_0^{\infty} \frac{dt}{t} \, \int \dd x
\sqrt{\gamma} \, \mathscr{G}_{ii} (x,x; t)\,,
}[oneLoopHeatKernel]
with $D$ the integer dimension of spacetime, justifying the assertion in
\zetaDeterminantRelation.  Its pole, which is the main interest to us,
may then be calculated with \hkansatz and by noting the coincident limits
therein, where $x' \rightarrow x$. If $D$ is even, then by expanding the
series in \hkansatz with $d = D - \epsilon$, it can be seen  that the only
piece that contains a pole in $\epsilon$ as $s \to 0$ is the $(D/2)$-th
piece. Hence, the object of concern for the pole of the effective action is
the coincident limit of the Seeley--DeWitt coefficient $a_{D/2,\lsp ii}$.
In six dimensions, in particular, we then have
\eqn{\left( - \tfrac{1}{2} \ln \det M \right)^{\text{pole}}_{\text{dim.} \,
\text{reg.}} = \frac{\mu^{-\epsilon}}{64\pi^3}\frac{1}{\epsilon}  \int \dd x
\sqrt{\gamma} \,[a_{3,\lsp ii}](x)\,,
}[oneLoopPoles]
where the $\mu^{-\epsilon}$ is inserted to preserve mass dimensions. The
coincident limit of $a_{3,\lsp ii} (x,x')$, denoted in \oneLoopPoles by the
brackets, may be found in appendix~\ref{appLimits}, equation
\eqref{aIIILimits}, and subsequently used to evaluate the one-loop
counterterms of $\tilde{S}^{(1)}$ of \OneLoopEffAction.

\begin{figure}[ht]
  \centering
  \begin{tikzpicture}
    \draw (0,0) circle [x radius=1cm, y radius=0.85cm];
    \draw (-1,0)--(1,0);
    \fill (-1,0) circle [radius=1pt] node[left] {$x$};
    \fill (1,0) circle [radius=1pt] node[right=0.5pt] {$x\smash{'}$};
  \end{tikzpicture}
  \hspace{1cm}
  \begin{tikzpicture}
    \draw (0,0) circle [radius=0.85cm];
    \filldraw[cross,fill=white] (-0.85,0) circle (4pt) node[left=2pt]
      {$x$};
  \end{tikzpicture}
  \caption{The diagrams that need to be considered at the two-loop level.}
  \label{fig:TwoLoopDiags}
\end{figure}
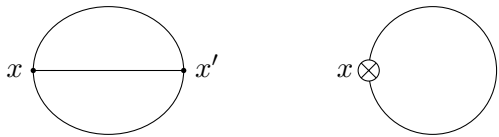

The task is then to extend the relatively graceful computation of the
one-loop effective action, {\it \`a la} \oneLoopHeatKernel, to higher loop
order. When using the heat kernel method in the context of the background
field method, two-loop and higher-order contributions to the effective
action are encompassed entirely within the calculation of vacuum bubble
diagrams. These are then evaluated in coordinate space by integrations over
the spacetime points involved in the loop diagram of the products of
Green's functions. It then becomes convenient, now specifying $d =
6-\epsilon$, to express the Green's function through the expansion
\hkansatz and \greensFxnHeatKernel which, after performing the integration,
yields
\eqna{G_{ij} (x,x') &= G_{0} (x,x') \,  a_{0,\lsp ij} (x,x') +
G_{1} (x,x') \,  a_{1,\lsp ij} (x,x')\\
&\quad+ R_2(x,x')\,a_{2,\lsp ij} (x,x') + R_3(x,x')\,a_{3,\lsp ij} (x,x')
+H (x,x')\,,}[propExpansion]
where the $H(x,x')$ term does not contribute to UV divergences of the
theory, i.e.\ do not have divergent behavior as $x' \to x$.\foot{It should
be noted, however, that these still have IR-divergent behavior. Although it
is not of interest in the calculation in this paper, it may be taken care
of by considering the log of the ratio of the determinant of interest,
equation \oneLoopHeatKernel, with determinant of the non-interacting
operator $M^{(0)}_{ij} = -\delta_{ij} \nabla^2$.  The ratio acts like a
normalized version of \oneLoopPoles and removes uninteresting IR
divergences. See~\cite[section 1]{Jack:1982hf} for details.} The utility of
this expansion is that it allows extraction of the poles of higher-loop
diagrams almost by inspection, once the $G_n(x,x')$ and $R_n (x,x')$ are
computed with \greensFxnHeatKernel. For example, in the two-loop case,
whose computation is detailed in section~\ref{polesEffAction}, the
coincident limit is necessary to evaluate the contribution there, so the
computation boils down to the coincident limits of the Seeley--DeWitt
coefficients, tabulated in the appendices, times coincident limits of the
$G_n(x,x')$ and $R_n(x,x')$, which are easily computed with knowledge of
the coincident limits of $\sigma (x,x')$ and $\Delta_{\text{VM}} (x,x')$,
also tabulated in the appendices.  Furthermore, although $G_{ij} (x,x')$
must be finite when $x \neq x'$ in $6$ or $6 - \epsilon$ dimensions,
products of the $G_n(x,x')$ and $R_n(x,x')$ may have poles in $\epsilon$.
In this two loop case the cubic product of $G_{ij}(x,x')$ is necessary, as
evinced in Fig.~\ref{fig:TwoLoopDiags} and~\eqref{twoLoopGraph}, and the
various products of the pieces of \propExpansion give rise to the poles
computed in section~\ref{polesEffAction}.

\subsec{Trace anomaly}
Now that the heat kernel method for the computation of the poles of the
effective action has been established, we can proceed to the computation of
the trace of the stress energy tensor defined by \lagrangian. To study the
trace it is useful to promote the metric and couplings to spacetime
dependent sources
\eqn{ \gamma^{\mu\nu} \rightarrow \gamma^{\mu\nu}(x)\,, \qquad g^I
\rightarrow g^I (x)\,, }[sources]
and subsequently promote the action of the theory to be diffeomorphism
invariant.  Then we can define the quantum stress-energy tensor and finite
composite operators by functionally differentiating with respect to these
sources:
\eqn{T_{\mu\nu}(x)=2\frac{\delta \tilde{S}_0}{\delta\gamma^{\mu\nu}(x)}\,,
\qquad [\cO_I(x)]=\frac{\delta \tilde{S}_0}{\delta g^I(x)}\,,}[]
where functional derivatives are defined in $d$ spacetime dimensions by
\eqn{\frac{\delta}{\delta\gamma^{\mu\nu}(x)}\gamma^{\kappa\lambda}(x')
=\delta_{\smash{(\mu}}^{\phantom{(\mu}\!\kappa}
\delta_{\smash{\nu)}}^{\phantom{\nu}\lambda}\,
\delta^{\hspace{0.5pt}d}(x,x')\,,\qquad
\frac{\delta}{\delta g^I(x)}g^J(x')=\delta_I^{\phantom{I}\!J}
\delta^{\hspace{0.5pt}d}(x,x')\,,}[]
with $X_{(I}Y_{J)}\equiv \tfrac12(X_IY_J+X_JY_I)$. With these definitions
it is easy to see that
\eqn{\gamma^{\mu\nu}T_{\mu\nu}=\epsilon\tilde{\mathscr{L}}_0+\nabla_\mu
I^\mu-(\Delta\phi)\cdot\frac{\delta}{\delta\phi}\tilde{S}_0 \,,}[TraceT]
where $\Delta$ is the canonical scaling dimension of $\phi$ and $I^\mu$ arises from
variations of curvature-dependent terms in $\tilde{\mathscr{L}}_0$.

The trace anomaly may be viewed as the theory's response to the local Weyl
rescalings
\eqn{\gamma^{\mu\nu}(x)\rightarrow \big(1+2\lsp\sigma (x)\big)
\lsp\gamma^{\mu\nu}(x)\,,
\qquad g^I(x)\rightarrow g^I(x)+\sigma (x)\lsp\hat{\beta}^I (x)\,.
}[weylVariations]
The scalar $\sigma(x)$ here is a variational parameter and should not be
confused with the biscalar geodetic interval $\sigma (x,x')$ of the
previous section.  At the level of the generating functional we can
implement these infinitesimal local Weyl transformations with the
generators
\eqn{\DeltaW_\sigma=2\int \dd x\sqrt{\gamma}\,\sigma\gamma^{\mu\nu}
\frac{\delta}{\delta\gamma^{\mu\nu}}\,,\qquad \Delta_\sigma^{\hat{\beta}}
=\int \dd x\sqrt{\gamma}\,\sigma\hat{\beta}^I \frac{\delta}{\delta
g^I}\,.}[weylGenerators]
With these definitions,
\eqn{\DeltaW_\sigma W=-\int \dd x\sqrt{\gamma}\,\sigma\gamma^{\mu\nu}
\langle T_{\mu\nu}\rangle\,,\qquad
\Delta_\sigma^{\hat{\beta}} W=-\int \dd x\sqrt{\gamma}\,\sigma\langle
\hat{\beta}^I[\cO_I]\rangle\,.}[]

Now, it is easy to see that the term $\epsilon\tilde{\mathscr{L}_0}$ in
\TraceT can be substituted with the use of \RGEFull, and so \TraceT can be
written in the form
\eqn{\gamma^{\mu\nu}T_{\mu\nu}-\hat{\beta}^I[\cO_I]\supset-\mu^{-\epsilon}(\beta_\lambda\cdot\mathscr{R}-\nabla_\mu
Z^\mu),}[TraceAnomT]
where $Z^\mu$ is the part of $I^\mu$ of \TraceT that contains
field-independent terms.\foot{Note that the field-dependent part of $I^\mu$
is responsible for $\cO_I\to[\cO_I]$, for the difference between
$\partial/\partial g^I$ and $\delta/\delta g^I$ is a total derivative.} In
\TraceAnomT we neglect field-dependent contributions besides those in
$\beta^I[\cO_I]$.  Equivalently, we can write
\eqn{\DeltaW_\sigma W-\Delta_\sigma^{\hat{\beta}}
W\supset\int \dd x\sqrt{\gamma}\,\sigma\mu^{-\epsilon}
\beta_\lambda \cdot\mathscr{R}+\int \dd x\sqrt{\gamma}\,
\partial_\mu\sigma\,\mu^{-\epsilon}Z^\mu.}[TraceAnomDelta]
Terms in the right-hand side of \TraceAnomT have been computed in
\cite{Jack:1990eb} for field theories in $d=4$. In this work we will
compute such terms for general multiflavor $\phi^3$ field theories in $d=6$. As we just
saw, these computations give results on the various terms that appear in
the consistency conditions derived from
\TraceAnomDelta~\cite{Grinstein:2013cka}.

Thus the relevant contributions to the trace of the stress energy tensor
have their origin in the $\lambda \cdot \mathscr{R}$ terms, which are, in
turn, obtained from the heat kernel methods of the previous section. The
$\beta_{\lambda} \cdot \mathscr{R}$ terms are computed from the $\lambda
\cdot \mathscr{R}$ terms by \RGEct, and the $Z^{\mu}$ terms are obtained
from the Weyl variation $\delta_\sigma(-\lambda \cdot \mathscr{R})$. One
can also change the basis so that terms in the variation
$\delta_\sigma(-\lambda \cdot \mathscr{R})$ that appear in $Z^{\mu}$ in one
basis appear in $\beta_\lambda\cdot\mathscr{R}$ in another and vice-versa.

\newsec{Weyl consistency conditions}[weylCCsSec]
The trace anomaly as presented in \TraceAnomDelta is useful because it
allows very powerful statements about the structure of the theory along the
renormalization group flow to be made. These statements arise from the Weyl
consistency conditions, a specific example of the Wess--Zumino consistency
conditions~\cite{Wess:1971yu} that constrain the form of a quantum anomaly
based upon the algebra of the anomalous symmetry group.

Consider the two generators acting on the connected diagram generating
functional $W$ in \TraceAnomDelta. We may take the commutator of their
actions on $W$ for two different variational parameters $\sigma$ and
$\sigma'$ and, because the Weyl rescalings are Abelian, obtain the Weyl consistency condition
\eqn{ \big[ \DeltaW_{\sigma} - \Delta^{\hat{\beta}}_{\sigma} , \,
\DeltaW_{\sigma'} - \Delta^{\hat{\beta}}_{\sigma'} \big] W = 0\,.
}[weylCCs]
Now, the terms in $W$, namely those coming from \Ltilde, have complicated
transformations under \weylGenerators and so \weylCCs imposes a set of
non-trivial constraints and relations among these terms.

In particular, as argued in~\cite{Osborn:1991gm}, $\lambda \cdot
\mathscr{R}$ must contain all terms that are diffeomorphism-invariant and,
by simple power-counting, of mass dimension $d$ that might arise in
addition to the usual operators in $\mathscr{L}$ in \Ltilde from the
promotion of the metric and couplings to spacetime-dependent sources as in
\sources. In two dimensions there are precisely three terms that fit the
bill, with a single resulting consistency constraint. In four dimensions
there are sixteen candidates, with seven independent consistency
conditions.  In six dimensions there are ninety five candidates with many
independent consistency constraint equations.

While the two- and four-dimensional cases are rather tractable and admit
relatively simple interpretations of the consistency conditions, the
six-dimensional case is significantly more complex. The $\beta_\lambda
\cdot \mathscr{R}$ and $Z^\mu$ terms as well as the consistency conditions in six
dimensions were categorized in~\cite{Grinstein:2013cka}. There, it was
established that the most interesting consistency condition found in two
and four dimensions, the consistency condition governing the flow of a
certain $a$-function along the renormalization group flow, survives in six
dimensions. The key point is to identify the analogs of the terms relevant
to this consistency condition from two and four dimensions. There they
involve the Euler density in the specific spacetime dimensionality and, in
four dimensions, terms involving the Einstein tensor. In six dimensions, in
fact, the coefficient $a$ of the six-dimensional Euler term $E_6$ is
related to terms involving the generalization of the Einstein tensor, the
so-called Lovelock tensor~\cite{Lovelock:1971yv}, in a way that is
almost\foot{In six dimensions there are curvature terms with coefficients
called ``vanishing anomalies'' whose only analog in lower dimensions is
the $R^2$ term in four dimensions. At fixed points they vanish (hence the
name) and the $a$ in six dimensions is completely analogous to lower
dimensions, however away from fixed points they modify $a$ to $\tilde{a}$
(to be discussed shortly) in a way distinct from two and four dimensions.}
completely analogous to the two- and four-dimensional settings.

To be clear, in six dimensions \TraceAnomDelta takes the form
\cite{Grinstein:2013cka}
\eqn{(\DeltaW_\sigma - \Delta_\sigma^\beta) W = \sum_{p=1}^{65}\int
d^{\hspace{0.5pt}6} x\sqrt{\gamma}
\,\sigma \left(\beta_{\lambda} \cdot \mathscr{R} \right)_p
+ \sum_{q=1}^{30}\int d^{\hspace{0.5pt}6} x \sqrt{\gamma}
\,\partial_\mu\sigma\,\mathscr{Z}^\mu_q\,.}[anomalyBasis]
The terms of interest to the aforementioned consistency condition are
contained therein
\eqna{(\DeltaW_\sigma - \Delta_\sigma^\beta) W &\supset \int
d^{\hspace{0.5pt}6}\hspace{-0.5pt}x\sqrt{\gamma}\,
\sigma\Big(\!-aE_6-b_1L_1-b_3L_3
+\tfrac12\cH_{IJ}^1\,\partial_\mu g^I\partial_\nu g^J\llsp
H^{\mu\nu}_1\Big)
\\
&\quad+ \int d^{\hspace{0.5pt}6}\hspace{-0.5pt}x\sqrt{\gamma}\,
\partial_\mu\sigma\,\cH_I^1\,\partial_\nu g^I\llsp H^{\mu\nu}_1\,,
}[relevantTerms]
where $L_{1,3}$, given in Appendix \ref{appConv}, are dimension-six
curvature terms whose coefficients $b_{1,3}$ vanish at fixed points,
$H_{1}^{\mu\nu}$ is the Lovelock curvature tensor, given in \eqref{BFour}, and the
coefficients $\cH_{IJ}^1$ and $\cH_I^1$ are functions of the coupling
constants.  By varying each piece in \relevantTerms with \weylGenerators
and applying the Weyl consistency conditions \weylCCs one obtains the
constraint equation
\eqn{\partial_I\tilde{a}=\tfrac16\cH_{IJ}^1\beta^J
+\tfrac16(\partial_{I}\cH_{J}^1-\partial_{J}\cH_{I}^1)\beta^J\,,}[aCC]
with
\eqn{\tilde{a}=a +\tfrac16 b_1-\tfrac{1}{90}b_3+\tfrac16\cH_I^1\beta^I\,.
}[atilde]
This equation is analogous to those found by Osborn in two and four
dimensions~\cite{Osborn:1991gm}.

By contracting with $\beta^I$ on each side of \aCC we arrive at
six-dimensional equivalent of Zamolodchikov's theorem from two
dimensions~\cite{Zamolodchikov:1986gt},
\eqn{\beta^I\partial_I\tilde{a}=\tfrac16\cH^1_{IJ}
\beta^I\beta^J\,.}[finalCC]
A similar relation was shown to hold in any even-dimensional spacetime
in~\cite{Grinstein:2013cka}. Thus, if we can compute $\cH_{IJ}^1$ in our
theory and establish a definite sign, the monotonicity of the
renormalization group flow of the theory can be established by way of
\finalCC.

All that remains now is to determine the terms in the consistency
conditions from the computation of the two-loop effective potential. The
following section sets itself to this task.

\newsec{Poles of the effective action}[polesEffAction]
In this section we present our results for the pole part of the effective
action up to two loops. For completeness we will also include here a mass
term in our theory as well as a term linear in $\phi$, i.e.\ we will take
our Lagrangian to be given by
\eqn{\mathscr{L} (\phi, g, m, h, \gamma) =\tfrac12\big(\partial_\mu\phi_i\,\partial_\nu\phi_i
\,\gamma^{\mu\nu} + (\xi_{ij} R+m_{ij}) \lsp\phi_i \phi_j\big) +h_i\phi_i
+ \tfrac{1}{3!} g_{ijk} \phi_i\phi_j\phi_k\,,}[lagrangianWithM]
where $m_{ij}$ and $h_i$ have mass-dimension two and four respectively.
Then, \Ltilde is modified to
\eqn{\tilde{\mathscr{L}}_0 = \mathscr{L}_0
-\mu^{-\epsilon}\lambda\cdot\mathscr{R}+\mu^{-\epsilon}
\mathscr{F} + \mu^{-\epsilon}\mathscr{M}\,,
}[LtildeWithM]
with $\mathscr{F} =
\mathscr{F}(\phi,g,\gamma)$ and $\mathscr{M} =
\mathscr{M}(\phi,g,m,\gamma)$. As in \Ltilde, in \LtildeWithM all
quantities are bare quantities that may be written in terms of the
renormalized quantities via \BareToRen.
As stated in section~\ref{secHeatKernel}, at one loop the effective action
is related to the Seeley--DeWitt coefficient $a_{D/2,\lsp ij}$ with $D$ the
even integer spacetime dimension.  To wit, in the theory
of~\lagrangianWithM in six dimensions, as in \oneLoopPoles,
\eqn{(- \tfrac{1}{2} \ln \det M)^{\text{pole}}
= \frac{1}{\epsilon} \frac{\mu^{-\epsilon}}{64\pi^3} \int \dd x
\sqrt{\gamma} \,  [a_{3,\lsp ii}] (x)\, .}[]
This result produces all terms in~\LtildeWithM.  Thus, at one loop, using
the result from appendix~\ref{appLimits}, discarding total derivatives, and
choosing $\xi_{ij}=\left( \tfrac15 - \tfrac{\epsilon}{100}
\right)\delta_{ij}$ from here on,\foot{The $\epsilon$-dependent portion is
necessary to maintain, in the absence of $m_{ij}$ and $h_i$, classical
conformal invariance of \lagrangianWithM in $6-\epsilon$ dimensions.
Although inconsequential at one loop, this $\epsilon$-dependence is crucial
at two loops. This has also been seen in higher-loop computations in four
dimensions in \cite{Brown:1980qq}. We thank Hugh Osborn for bringing this
issue to our attention.} we can isolate each piece of the one-loop
effective action according to~\LtildeWithM and~\BareToRen. We
find\foot{This agrees with the results of~\cite{Bastianelli:2000hi}, where
similar computations were performed for fermions and two-forms.}
\eqn{\lambda^{(1)}\cdot\mathscr{R}=\tfrac{1}{\epsilon}\tfrac{1}{64\pi^3}
n_\phi\left(-\tfrac{1}{9072}\lsp E_6+\tfrac{1}{540}\lsp I_1
-\tfrac{1}{3024}\lsp I_2-\tfrac{1}{2520}\lsp I_3\right),}[lambdaOne]
where $n_\phi$ is the number of scalar fields $\phi$. The field-dependent
counterterms at one loop are given by
\eqn{Z_{ij}^{(1)}=-\tfrac{1}{\epsilon}\tfrac{1}{64\pi^3}\tfrac{1}{6}\lsp
g_{ikl}g_{jkl}\,,\qquad
L_{ijk}^{(1)}=-\tfrac{1}{\epsilon}\tfrac{1}{64\pi^3}\big(
g_{imn}g_{jmp}g_{knp}-\tfrac{1}{12}(g_{ijl}g_{kmn}g_{lmn}
+\text{permutations})\big)\,,}[ZLOne]
and
\eqn{\mathscr{F}^{(1)}=-\tfrac{1}{\epsilon}\tfrac{1}{64\pi^3}\tfrac16
\big(\tfrac{1}{30}\lsp Fg_{ijj}\phi_i+\tfrac{1}{10}\lsp R\lsp
g_{ikl}g_{jkl}\lsp \phi_i\phi_j
+\tfrac{1}{2}\lsp\partial^\mu g_{ikl}\lsp\partial_\mu g_{jkl}
\lsp\phi_i\phi_j + g_{ikl}\lsp\partial^\mu
g_{jkl}\lsp\partial_\mu\phi_i\,\phi_j\big)\,.}[FOne]
Finally, the mass-dependent counterterms are
\eqna{\mathscr{M}^{(1)}=-\tfrac{1}{\epsilon}\tfrac{1}{64\pi^3}\tfrac12
\big(&\tfrac{1}{90}\lsp F\lsp m_{ii}+\tfrac{1}{30}\lsp R\lsp m_{ij}m_{ij}\\
&+\tfrac{1}{15}((R-5\lsp\nabla^2)m_{ij})\lsp g_{ijk}\lsp\phi_k
+m_{ij}\lsp g_{ikl}g_{jkm}\lsp\phi_l\phi_m
+m_{ij}m_{ik}\lsp g_{jkl}\lsp\phi_l\\
&+\tfrac{1}{6}\lsp\partial^\mu m_{ij}\lsp\partial_\mu m_{ij}
+\tfrac13\lsp m_{ij}m_{jk}m_{ki}\big)\,.}[MOne]

At two loops in the background field method we must compute the relevant
vacuum bubble diagrams.  Thus we are led to consider the diagrams in
Fig.~\ref{fig:TwoLoopDiags}, where in the diagram on the right
\tikz{\filldraw[cross,fill=white] (0,0) circle (4pt);} denotes the one-loop
counterterm.
Note that these are graphs in position space, and that short distance
singularities arise here from the coincident limit of products of
position-space propagators. In particular, the left graph in
Fig.~\ref{fig:TwoLoopDiags} is given by
\eqn{
\tikz[baseline=(vert_cent.base)]{
  \node (vert_cent) {\hspace{-13pt}$\phantom{-}$};
  \draw (0.0,0) ++(0:0.5cm) arc (0:360:0.5cm and 0.4cm)
        (-0.5,0.0)--(0.5,0.0);
} \lsp=\tfrac{1}{12}\mu^\epsilon
	\int \, \dd x \, \dd x' \sqrt{\gamma} \, \sqrt{\gamma'} \,
 g_{ijk}(x)\lsp g_{lmn} (x') \,
G_{il} (x,x') \, G_{jm} (x,x') \, G_{kn}
(x,x')\,, }[twoLoopGraph]
with the factor of $\tfrac{1}{12}$ a symmetry factor from interchanging the
Green's functions.  The evaluation of this integral is straightforward,
once the divergent parts of the products of the $G_n(x,x')$ from
\propExpansion are known; as these are listed in appendix~\ref{appProp}, we
will not explicitly show the intermediate details of the cube of the
propagator in \twoLoopGraph and simply include their contributions to the
counterterms in \LtildeWithM in the results listed below.

The graph on the right in Fig.~\ref{fig:TwoLoopDiags} is slightly more
complicated than the expression listed in~\cite{Jack:1985wf} because of the
spacetime dependence of the couplings. The counterterm insertion to the
propagator in that diagram is simply obtained from our one-loop results.
More specifically, using our results \FOne and \MOne, or, equivalently, the
coincident limit of $a_{3,\lsp ii}(x,x')$ in \eqref{aIIILimits}, we obtain
\begin{align}
\tikz[baseline=(vert_cent.base)]{
    \draw (0,0) circle [radius=0.55cm];
    \filldraw[cross,fill=white] (-0.55,0) circle (4pt) node[left=2pt]
      {};
} =
	\frac{1}{\epsilon}\frac{1}{64\pi^3}
	\int \, \dd x \, \dd x' \, \sqrt{\gamma} \,\sqrt{\gamma'} \,
	\delta^{\llsp d} (x,x') \,
	\Big[
	& -\tfrac{1}{2}\left(
	g_{lmn} \phi_n + m_{lm} + (\xi_{lm} - \tfrac{1}{6} \delta_{lm})R \right)\lsp \nonumber \\
	& +\tfrac{1}{12} \delta_{lm} \,
\partial_x^\mu(\partial_{x'})_\mu \Big] \big(g_{ikl}(x)g_{jkm}(x')\lsp
	G_{ij}(x,x')\big)  \,.
\label{eq:twoLoopCT}
\end{align}
%[twoLoopCT]
%
Here we have specified the spacetime point at which any derivatives are to
be taken.

The UV divergences of these graphs are obtained from the divergences that
arise from the products of propagators, discussed at the end of
section~\secHeatKernel. We use the results listed in
appendix~\ref{appProp}, along with prudent integrations by parts, to reduce
both \twoLoopGraph and~(\ref{eq:twoLoopCT}) to poles in $\epsilon$ and
coincident limits of the Seeley--DeWitt coefficients, which ultimately give
explicit expression for the $-\mu^{-\epsilon}\lambda\cdot\mathscr{R}$,
$\mu^{-\epsilon} \mathscr{F}(\phi)$, and $\mu^{-\epsilon}\mathscr{M}(m)$
pieces of \LtildeWithM. In particular, for the counterterm graph, we
find\foot{This equation may be compared with (3.14b)
of~\cite{Jack:1985wf}.  Note that our result reduces to this equation in
the single spacetime-independent coupling case.}
\eqna{
\tikz[baseline=(vert_cent.base)]{
    \draw (0,0) circle [radius=0.55cm];
    \filldraw[cross,fill=white] (-0.55,0) circle (4pt) node[left=2pt]
      {};
} =
\frac{1}{\epsilon^2}\frac{\mu^{-\epsilon}}{(64\pi^3)^2}
	 \int \, \dd x  \sqrt{\gamma} \,
	 \Big[
		\Big(
		&
		-\tfrac{1}{2}\lsp g_{ikl}g_{jkm}
		\left(
		g_{lmn} \phi_n + m_{lm} + \left( \xi_{lm} - \tfrac{7}{36} \delta_{lm}
		\right)R \right)
		\\
		&
		- \tfrac{1}{12}\lsp\partial^{\mu} g_{ikl}\lsp \partial_{\mu} g_{jkl}
		\Big)\cdot 2\lsp \big[a_{2,\lsp ij}\big] (x)
		\\
		&
		+ \tfrac{1}{12}\lsp g_{ikl} g_{jkl} \left( 2\big[\nabla^2
			a_{2,\lsp ij}\big](x) - (6-\epsilon) \big[a_{3,\lsp ij}\big](x)
            \right)\Big]\,.}[poleEqnB]

Evaluting these graphs with the coincident limits listed in
appendices~\ref{appLimits} and~\ref{appProp}, we generate the terms as
listed in~\LtildeWithM at the two loop level.  For the curvature and
derivatives on couplings terms, at two loops discarding total derivatives
we have
\eqna{\lambda^{(2)}\cdot\mathscr{R}=\tfrac{1}{\epsilon}\tfrac{1}
{(64\pi^3)^2}
&\tfrac{1}{1620}
\big(-\tfrac{1}{36}\lsp g_{ijk}g_{ijk}\left(I_1
-\tfrac{13}{4}\lsp I_2-\tfrac{9}{8}\lsp I_3\right)\\
&+\tfrac{3}{16}\big(\partial^\mu\nabla^2g_{ijk}
\,\partial_\mu\nabla^2g_{ijk}-4\lsp R^{\mu\nu}\lsp\nabla_\mu\partial_\nu
g_{ijk}\lsp\nabla^2g_{ijk}+R\lsp\nabla^2g_{ijk}\lsp\nabla^2g_{ijk}\big)\\
&-\tfrac{1}{8}\big(H_1^{\mu\nu}+30\lsp H_2^{\mu\nu}
-8\lsp H_3^{\mu\nu}-10\lsp H_4^{\mu\nu}
-3\lsp H_5^{\mu\nu}\big)\partial_\mu g_{ijk}\lsp\partial_\nu g_{ijk}\\
&+\left(E_4 +\tfrac{49}{480}\lsp F+\tfrac{7}{400}\lsp R^2
-\tfrac{9}{80}\lsp\nabla^2R\right)\partial^\mu g_{ijk}\lsp
\partial_\mu g_{ijk}\big)\,.}[lambdaTwo]
This result has been given in a conformally-covariant basis in
\cite{Osborn:2015rna}. Field-dependent counterterms at two loops are given
by
\eqn{Z_{ij}^{(2)}=-\tfrac{1}{\epsilon}\tfrac{1}{(64\pi^3)^2}\tfrac{1}{18}
\big(g_{ikl}g_{jmn}g_{kmp}g_{lnp}
-\tfrac{11}{24}\lsp g_{ikl}g_{jkm}g_{lnp}g_{mnp}\big)\,,}[ZTwo]
\eqna{L_{ijk}^{(2)}&=-\tfrac{1}{\epsilon}\tfrac{1}{(64\pi^3)^2}\big(
\tfrac14(g_{ilm}g_{jnp}g_{kqr}g_{lnq}g_{mpr}
+\tfrac32\lsp g_{ilm}g_{jln}g_{kpq}g_{mpr}g_{nqr}
-\tfrac{7}{12}\lsp g_{ilm}g_{jln}g_{kmp}g_{nqr}g_{pqr})\\
&\hspace{2.3cm}-\tfrac{1}{36}(g_{ijl}g_{kmn}g_{lpq}g_{mpr}g_{nqr}
-\tfrac{11}{24}g_{ijl}g_{kmn}g_{lmp}g_{nqr}g_{pqr}
+\text{permutations})\big)\,,}[LTwo]
which are relevant for the two-loop anomalous dimension and beta function,
and
\begin{align}
\mathscr{F}^{(2)}=-\tfrac{1}{\epsilon}\tfrac{1}{(64\pi^3)^2}
\tfrac{1}{18}\Big(\hspace{-2pt}&\big(\tfrac{47}{2880}\lsp F\lsp
g_{ijk}g_{klm}g_{jlm}-
\tfrac{13}{48}\lsp R^{\mu\nu}\lsp g_{ijk}\lsp \partial_\mu g_{jmn}\lsp
\partial_\nu g_{kmn}\nonumber\\
&\qquad+\tfrac{1}{80}\lsp R\lsp g_{ijk}g_{jlm}\nabla^2 g_{klm}
+\tfrac{101}{2400}\lsp g_{ijk}\lsp \partial^\mu g_{jlm}\lsp
\partial_{\mu}g_{klm}\nonumber\\
&\qquad-\tfrac{3}{32}\lsp g_{ijk}\lsp\nabla^2 g_{jlm}\lsp
\nabla^2 g_{klm}
-\tfrac{7}{48}\lsp g_{ijk}\nabla^\mu\partial^\nu g_{jlm}\lsp
\nabla_\mu\partial_\nu g_{klm}\nonumber\\
&\qquad-\tfrac{19}{48}\lsp g_{ijk}\lsp\partial^\mu\nabla^2 g_{jlm}
\lsp\partial_\mu g_{klm}
-\tfrac{1}{16}\lsp g_{ijk}g_{jlm}\lsp \nabla^2\nabla^2 g_{klm}\big)\phi_i\nonumber\\
&+\big(\tfrac{23}{200}\lsp R\lsp g_{ikl}g_{jmn}g_{kmp}g_{lnp}-
\tfrac{29}{600}\lsp R\lsp g_{ikl}g_{jkm}g_{lnp}g_{mnp}\nonumber\\
&\qquad+\tfrac12\lsp g_{klm}g_{knp}\lsp\partial^\mu g_{iln}\lsp \partial_\mu
g_{jmp}
-\tfrac{11}{48}\lsp g_{klm}g_{kln}\lsp\partial^\mu g_{imp}\lsp\partial_\mu
g_{jnp}\nonumber\\
&\qquad-\tfrac12\lsp g_{ikl}g_{kmn}\lsp\partial^\mu g_{jlp}\lsp
\partial_\mu g_{mnp}
+\tfrac74\lsp g_{ikl}g_{kmn}\lsp\partial^\mu g_{jmp}\lsp
\partial_\mu g_{lnp}\nonumber\\
&\qquad-\tfrac{11}{24}\lsp g_{ikl}g_{mnp}\lsp\partial^\mu g_{jkm}\lsp
\partial_\mu g_{lnp}
+\tfrac34\lsp g_{ikl}g_{jmn}\lsp \partial^\mu g_{kmp}\lsp
\partial_\mu g_{lnp}\nonumber\\
&\qquad\hspace{4.5cm}-\tfrac{1}{16}\lsp g_{ikl}g_{jkm}\lsp\partial^\mu
g_{lnp}\lsp \partial_\mu g_{mnp}\big)\phi_i\phi_j\nonumber\\
&+\big(g_{jkl}g_{kmn}g_{lmp}\lsp\partial^\mu
g_{inp}+\tfrac74\lsp g_{ikl}g_{jmn}g_{kmp}\lsp
\partial^\mu g_{lnp}\nonumber\\
&\qquad-\tfrac{11}{24}\lsp g_{jkm}g_{mnp}\lsp\partial^\mu (g_{ikl}g_{lnp})
-\tfrac12\lsp g_{ikl}g_{jkm}g_{lnp}\lsp\partial^\mu g_{mnp}\big)
\phi_i\lsp\partial_\mu\phi_j\Big)\,.
\label{eq:FTwo}
\end{align}
%[FTwo]
%
The mass-dependent counterterms are given by
\begin{align}
\mathscr{M}^{(2)}=-\tfrac{1}{\epsilon}\tfrac{1}{(64\pi^3)^2}
\tfrac{1}{18}&\Big(\hspace{-2pt}\big(\tfrac{47}{2880}\lsp F\lsp g_{ikl}
g_{jkl}
-\tfrac{13}{48}\lsp R^{\mu\nu}\lsp\partial_\mu g_{ikl}
\lsp\partial_\nu g_{jkl}\nonumber\\
&\qquad+\tfrac{1}{80}\lsp R\lsp g_{ikl}\lsp\nabla^2 g_{jkl}
+\tfrac{101}{2400}\lsp R\lsp\partial^\mu g_{ikl}\lsp \partial_\mu
g_{jkl}\nonumber\\
&\qquad-\tfrac{3}{32}\lsp \nabla^2 g_{ikl}\nabla^2 g_{jkl}
-\tfrac{7}{48}\lsp \nabla^\mu\partial^\nu g_{ikl}\lsp \nabla_\mu\partial_\nu
g_{jkl}
-\tfrac13\lsp\partial^\mu g_{ikl}\lsp\partial_\mu\nabla^2 g_{jkl}\nonumber\\
&\qquad+\tfrac{1}{10}\lsp R^{\mu\nu}\lsp g_{ikl}\lsp\partial_\mu
g_{jkl}\lsp\partial_\nu
-\tfrac{1}{20}\lsp R\lsp g_{ikl}\lsp\partial^\mu g_{jkl}\lsp \partial_\mu
+\tfrac{1}{16}\lsp g_{ikl}\lsp\partial^\mu\nabla^2 g_{jkl}\lsp \partial_\mu
\nonumber\\
&\qquad+\tfrac{23}{100}\lsp R\lsp g_{ikl}g_{jkm} g_{lmn}\lsp
\phi_n
-\tfrac{29}{300}\lsp R\lsp g_{ikl}g_{jmn}g_{kmn}\lsp \phi_l\nonumber\\
&\qquad-\tfrac74\lsp g_{ikl}\nabla^2 g_{jkm}\lsp g_{lmn}\lsp\phi_n
+\tfrac{11}{24}\lsp g_{ikl}\nabla^2 g_{jmn}\lsp g_{kmn}\lsp\phi_l\nonumber\\
&\qquad\hspace{5cm}+\tfrac12\lsp g_{ikl}g_{jmn}\nabla^2 g_{kmn}\lsp
\phi_l\nonumber\\
&\qquad-\tfrac14\lsp\partial^\mu g_{ikl}\lsp\partial_\mu
g_{jkm}\lsp g_{lmn}\lsp\phi_n
+\tfrac56\lsp g_{ikl}\lsp\partial^\mu g_{jmn}\lsp \partial_\mu
g_{mnk}\lsp\phi_l\nonumber\\
&\qquad-2\lsp g_{ikl}\partial^\mu g_{jkm}\lsp g_{lmn}\lsp\phi_n\lsp
\partial_\mu
+\tfrac{5}{12}\lsp g_{ikl}\lsp\partial^\mu g_{jmn}\lsp
g_{kmn}\lsp\phi_l\lsp\partial_\mu\nonumber\\
&\qquad\hspace{5cm}+\tfrac12\lsp g_{ikl}g_{jmn}\lsp\partial^\mu g_{kmn}
\lsp\phi_l\lsp\partial_\mu\nonumber\\
&\qquad-g_{ikl}g_{jkm}g_{lmn}\lsp\phi_n\lsp\nabla^2
+\tfrac{11}{24}\lsp g_{ikl}g_{jmn}g_{kmn}\lsp \phi_l\lsp\nabla^2\nonumber\\
&\qquad+\tfrac94\lsp g_{ikl}g_{jmn}g_{kmp}g_{lnq}\lsp\phi_p\phi_q
+\tfrac98\lsp g_{ikl}g_{jkn}g_{lpq}g_{npr}\lsp\phi_q\phi_r\nonumber\\
&\qquad+\tfrac94\lsp g_{ikl}g_{jmn}g_{kmp}g_{npq}\lsp\phi_l\phi_q
-\tfrac78\lsp g_{ikl}g_{jmn}g_{kpq}g_{mnp}\lsp\phi_l\phi_q\nonumber\\
&\qquad\hspace{5cm}-\tfrac{7}{16}\lsp g_{ikl}g_{jmn}g_{kpq}
g_{mpq}\lsp\phi_l\phi_n \big)m_{ij}\nonumber\\
&+\big(\tfrac{23}{200}\lsp R\lsp g_{ikm}g_{jlm}
-\tfrac{29}{600}\lsp R\lsp g_{imn}g_{kmn}\lsp\delta_{jl}\nonumber\\
&\qquad+\tfrac34\lsp\partial^\mu g_{ikm}\lsp\partial_\mu g_{jlm}
-\tfrac{1}{16}\lsp\partial^\mu g_{imn}\lsp\partial_\mu g_{kmn}\lsp
\delta_{jl}\nonumber\\
&\qquad+\tfrac94\lsp g_{imn}g_{jkp}g_{lmp}\lsp\phi_n
+ \tfrac94 \lsp g_{ikm}g_{jln}g_{mnp}\lsp\phi_p
-\tfrac{7}{16}\lsp g_{ikm}g_{jnp}g_{lnp}\lsp\phi_m\nonumber\\
&\qquad+\tfrac98\lsp g_{imn}g_{kmp}g_{npq}\lsp\phi_q
\lsp\delta_{jl}
-\tfrac{7}{8}\lsp g_{imn}g_{kpq}g_{mnp}\lsp\phi_q\lsp
\delta_{jl}\big)m_{ij}m_{kl}\nonumber\\
&+\big(\tfrac{7}{4}\lsp g_{ikm}\lsp\partial^\mu g_{jlm}\lsp
-\tfrac12\lsp g_{imn}\lsp\partial^\mu g_{kmn}\delta_{jl}
-\tfrac{11}{24}\lsp \partial^\mu g_{imn}\lsp g_{kmn}\lsp\delta_{jl}\big)
m_{ij}\lsp\partial_\mu m_{kl}\nonumber\\
&+\big(\tfrac12\lsp g_{ikm}g_{jlm}
-\tfrac{11}{48}\lsp g_{imn}g_{kmn}\lsp\delta_{jl}\big)
\partial^\mu m_{ij}\lsp\partial_\mu m_{kl}\nonumber\\
&+\big(\tfrac34\lsp g_{ikm}g_{jln}
+\tfrac98\lsp g_{ikp}g_{lmp}\lsp\delta_{jn}
-\tfrac{7}{16}\lsp g_{ipq}g_{kpq}\lsp
\delta_{jm}\delta_{ln}\big)m_{ij}m_{kl}m_{mn}\Big)\,.
\label{eq:MTwo}
\end{align}
%[MTwo]
%

Although these terms are unsightly, they allow us to calculate the
quantities of interest---they give us the complete, general trace anomaly
on a curved background with spacetime dependent marginal sources ($g_{ijk}
(x)$ and $\gamma_{\mu\nu}(x)$) via equation \TraceAnomT. Each of the terms
presented in this section yields the relevant beta functions in the first
part of \TraceAnomT via equation \RGEct. The second set of terms in
\TraceAnomT, called $\nabla_{\mu} Z^{\mu}$, are obtained from a Weyl
variation of the $\lambda \cdot \mathscr{R}$ terms, as seen in
\TraceAnomDelta. Since the Weyl variation is non-trivial, we report the
one- and two-loop contribution to $Z^{\mu}$ here, since it is required for
the identification of terms necessary to the computation of $\tilde{a}$ in
section~\ref{metricSec}.

At the one-loop level there are no contributions from the Weyl variation of
\lambdaOne and so $Z^{(1)\lsp\mu} = 0$. At two loops the Weyl variation of
\lambdaTwo yields
\begin{equation}
\begin{aligned}
	Z^{(2) \lsp \mu} = \tfrac{1}{(64\pi^3)^2} \tfrac{1}{1620}
	\Big(
	&
	E_4\lsp g_{ijk}\lsp \partial^{\mu} g_{ijk}
	+ \tfrac{49}{480}\lsp F\lsp g_{ijk}\lsp \partial^{\mu} g_{ijk}
	+ \tfrac{7}{400}\lsp R^2\lsp g_{ijk}\lsp \partial^{\mu} g_{ijk}
    - \tfrac{9}{80}\lsp \nabla^2 R\lsp g_{ijk}\lsp \partial^{\mu} g_{ijk}
    \\
	&
	- \tfrac{1}{8}\lsp H_1^{\mu\nu} g_{ijk}\lsp \partial_{\nu} g_{ijk}
	- \tfrac{15}{4}\lsp H_2^{\mu\nu} g_{ijk}\lsp \partial_{\nu} g_{ijk}
	+ H_3^{\mu\nu} g_{ijk}\lsp \partial_{\nu} g_{ijk}
	\\
	&
	\hspace{5cm}
	+ \tfrac{5}{4}\lsp H_4^{\mu\nu} g_{ijk}\lsp \partial_{\nu} g_{ijk}
	+ \tfrac{3}{8}\lsp H_5^{\mu\nu} g_{ijk}\lsp \partial_{\nu} g_{ijk}
	\\
	&
	+ \tfrac{3}{8}\lsp\nabla^{\mu} R^{\nu \rho}\lsp g_{ijk}\lsp \nabla_{\nu} \partial_{\rho} g_{ijk}
	+ \tfrac{47}{40}\lsp\nabla^{\mu} R^{\nu \rho}\lsp \partial_{\nu} g_{ijk} \lsp\partial_{\rho} g_{ijk}
	- \tfrac{1}{4}\lsp\nabla^{\nu} R^{\mu \rho}\lsp \partial_{\nu} g_{ijk} \lsp\partial_{\rho} g_{ijk}
	\\
	&
	\hspace{5cm}
	+ \tfrac{29}{40}\lsp\partial^{\nu} R\lsp\lsp \partial^{\mu} g_{ijk}\lsp \partial_{\nu} g_{ijk}
	+ \tfrac{43}{400}\lsp\partial^{\mu} R\lsp\lsp \partial_{\nu} g_{ijk} \lsp\partial^{\nu} g_{ijk}
	\\
	&
	+ \tfrac{3}{8}\lsp R^{\nu\rho}\lsp g_{ijk}\lsp \nabla^{\mu} \nabla_{\nu} \partial_{\rho} g_{ijk}
	+ \tfrac{3}{8}\lsp R^{\mu\nu}\lsp g_{ijk}\lsp \partial_{\nu} \nabla^2 g_{ijk}
	- \tfrac{3}{8}\lsp R^{\nu\rho}\lsp \partial^{\mu} g_{ijk}\lsp \nabla_{\nu} \partial_{\rho} g_{ijk}
	\\
	&
	- \tfrac{9}{8}\lsp  R^{\mu\nu} \partial_{\nu} g_{ijk}\lsp \nabla^2 g_{ijk}
	+ 4\lsp R^{\mu\rho} \partial^{\nu} g_{ijk}\lsp \nabla_{\nu}\partial_{\rho} g_{ijk}
	+ \tfrac{13}{5}\lsp R^{\nu\rho} \partial_{\nu} g_{ijk}\lsp \nabla^{\mu} \partial_{\rho} g_{ijk}
	\\
	&
    - \tfrac{3}{16}\lsp R\lsp g_{ijk}\lsp \partial^{\mu} \nabla^2 g_{ijk}
	+ \tfrac{73}{80}\lsp R\lsp \partial^{\mu} g_{ijk}\lsp \nabla^2 g_{ijk}
	- \tfrac{121}{100}\lsp R\lsp \nabla^{\mu} \partial_{\nu} g_{ijk}\lsp \partial^{\nu} g_{ijk}
	\\
	&
	+\tfrac{1}{2}\lsp R^{\mu\nu\rho\sigma} \partial_{\rho} g_{ijk}\lsp \nabla_{\nu} \partial_{\sigma} g_{ijk}
	\\
	&
	- \tfrac{3}{20}\lsp \nabla_{\nu} \partial_{\rho} g_{ijk}\lsp \nabla^{\mu}
	\nabla^{\nu} \partial^{\rho} g_{ijk}
	+ \tfrac{3}{10}\lsp \nabla^{\mu} \partial^{\nu} g_{ijk}\lsp\partial_{\nu}
	\nabla^2 g_{ijk}
	+ \tfrac{3}{16}\lsp \nabla^2 g_{ijk}\lsp \partial^{\mu} \nabla^2 g_{ijk}
	\\
	&
	- \tfrac{9}{20}\lsp \partial_{\nu} g_{ijk}\lsp \nabla^{\mu}\partial^{\nu}
	\nabla^2 g_{ijk}
	+ \tfrac{3}{16}\lsp \partial^{\mu} g_{ijk}\lsp \nabla^2\nabla^2 g_{ijk}
	\\
	&
	+ \tfrac{3}{16}\lsp g_{ijk}\lsp \partial^{\mu} \nabla^2\nabla^2 g_{ijk}
	\Big)
	\, .
\end{aligned}
\label{eq:Zmu}
\end{equation}
%[Zmu]
%
It should be noted that the basis reported here is not identical to the
$\mathscr{Z}^{\mu}_q$ terms reported as in \anomalyBasis, which refers to
the basis used in~\cite{Grinstein:2013cka} written by some of the authors
of the present work. However, as that basis is complete, the terms
in~(\ref{eq:Zmu}) may be written in the $\mathscr{Z}^{\mu}_q$ basis with
repeated and judicious integrations by parts. For the purposes of the
calculations in section~\ref{metricSec} of this paper, we did not find the
basis referred to in \anomalyBasis useful and were able to identify those
terms required in equations \eqref{DHFour} and \eqref{DHThree} from the
current presentation in~(\ref{eq:Zmu}).

From equations \lambdaOne to \MOne and \lambdaTwo to~(\ref{eq:MTwo})  we can extract
the beta functions for the couplings and masses, the anomalous dimensions
of the fields, and, perhaps most importantly, the quantity $a$ (and
$\tilde{a}$), as described in section~\ref{weylCCsSec}, which is the analog
of Zamolodchikov's celebrated $c$.

\newsec{Beta functions and anomalous dimensions}[betagammasec]
As we have seen using background field and heat kernel methods the
computation of $Z_{ij}$ and $L_{ijk}$ is easily done in position space and
does not require the calculation of any integrals. With our results \ZLOne,
\ZTwo and \LTwo we can now compute the anomalous dimension $\gamma_{ij}$ of
$\phi_i$ and the beta function $\beta_{ijk}$ to two-loop order.

The anomalous dimension is defined by
\eqn{\gamma=-Z^{-1/2}\frac{dZ^{1/2}}{dt}\,,\qquad
t=-\ln(\mu/\mu_0)\,,}[]
where the RG time $t$ is defined to increase as we flow to the IR. At one
loop we find
\eqn{\gamma^{(1)}=\frac{1}{64\pi^3}\frac{1}{12}
\tikz[baseline=(vert_cent.base)]{
  \node (vert_cent) {\hspace{-13pt}$\phantom{-}$};
  \draw (0,0)--(0.3,0)
        (0.7,0) ++(0:0.4cm) arc (0:360:0.4cm and 0.3cm)
        (1.1,0)--(1.4,0);
}\,,
}[gammaone]
where we use the diagram to denote the corresponding contraction of the
couplings, i.e.\
\eqn{
\tikz[baseline=(vert_cent.base)]{
  \node (vert_cent) {\hspace{-13pt}$\phantom{-}$};
  \draw (0,0)--(0.3,0)
        (0.7,0) ++(0:0.4cm) arc (0:360:0.4cm and 0.3cm)
        (1.1,0)--(1.4,0);
}
=g_{ikl}g_{jkl}\,.
}[]
The two-loop anomalous dimension is
\eqn{\gamma^{(2)}=\frac{1}{(64\pi^3)^2}\frac{1}{18}\left(
\tikz[baseline=(vert_cent.base)]{
  \node (vert_cent) {\hspace{-13pt}$\phantom{-}$};
  \draw (0,0)--(0.3,0)
        (0.7,0) ++(0:0.4cm) arc (0:360:0.4cm and 0.3cm)
        (0.7,0.3)--(0.7,-0.3)
        (1.1,0)--(1.4,0);
}
-\frac{11}{24}
\tikz[baseline=(vert_cent.base)]{
  \node (vert_cent) {\hspace{-13pt}$\phantom{-}$};
  \draw (0,0)--(0.3,0)
        (0.7,0) ++(0:0.4cm and 0.3cm) arc (0:50:0.4cm and 0.3cm) node (n1)
        {}
        (0.7,0) ++(50:0.4cm and 0.3cm) arc (50:130:0.4cm and 0.3cm) node
        (n2) {}
        (0.7,0) ++(130:0.4cm and 0.3cm) arc (130:360:0.4cm and 0.3cm)
        (n1.base) to[out=215,in=325] (n2.base)
        (1.1,0)--(1.4,0);
}
\right)\,.}[gammatwo]
For the case of a single field $\phi$ our results \gammaone and \gammatwo
reduce to the results of~\cite{Macfarlane:1974vp} (see
also~\cite{Toms:1982af, Kodaira:1985vr, Kodaira:1985pg, Jack:1985wf}).

The beta function is defined by
\eqn{\beta(g) = \mu\frac{dg}{d\mu}=-\frac{dg}{dt}\,.}[]
At one loop we find
\eqn{\beta^{(1)}=-\frac{1}{64\pi^3}\left(
\tikz[baseline=(vert_cent.base)]{
  \node (vert_cent) {\hspace{-13pt}$\phantom{-}$};
  \draw (0.7,0) ++(0:0.35cm) arc (0:135:0.35cm) node (n1) {}
        (0.7,0) ++(135:0.35cm) arc (135:225:0.35cm) node (n2) {}
        (0.7,0) ++(225:0.35cm) arc (225:360:0.35cm) node (n3) {}
        (n1.base)--+(145:0.5cm)
        (n2.base)--+(215:0.5cm)
        (n3.base)--+(0:0.5cm);
}
-\frac{1}{12}
\tikz[baseline=(vert_cent.base)]{
  \node (vert_cent) {\hspace{-13pt}$\phantom{-}$};
  \draw (0,0.5)--(0.5,0)
        (0,-0.5)--(0.5,0)
        (0.5,0)--(0.8,0)
        (1.2,0) ++(0:0.4cm) arc (0:360:0.4cm and 0.3cm)
        (1.6,0)--(1.9,0);
}
\right)\,,}[betaone]
where permutations of the free indices in the wavefunction-renormalization
correction are understood, i.e.\
\eqn{
\tikz[baseline=(vert_cent.base)]{
  \node (vert_cent) {\hspace{-13pt}$\phantom{-}$};
  \draw (0,0.5)--(0.5,0)
        (0,-0.5)--(0.5,0)
        (0.5,0)--(0.8,0)
        (1.2,0) ++(0:0.4cm) arc (0:360:0.4cm and 0.3cm)
        (1.6,0)--(1.9,0);
}
=g_{ijl}g_{lmn}g_{kmn}+\text{permutations}\,.}[]
Eq.~\betaone reproduces the result of~\cite{Macfarlane:1974vp} (see
also~\cite{Toms:1982af, Kodaira:1985vr, Kodaira:1985pg, Jack:1985wf}) in
the case of a single field $\phi$. In that case $\beta^{(1)}$ has a
negative sign, and hence the corresponding theory is asymptotically-free.
The two-loop beta function is
\eqn{\beta^{(2)}=-\frac{1}{(64\pi^3)^2}\frac12\left(
\tikz[baseline=(vert_cent.base)]{
  \node (vert_cent) {\hspace{-13pt}$\phantom{-}$};
  \draw (0.7,0) ++(0:0.35cm) arc (0:45:0.35cm) node (n1) {}
        (0.7,0) ++(45:0.35cm) arc (45:135:0.35cm) node (n2) {};
  \draw[white] (0.7,0) ++(135:0.35cm) arc (135:225:0.35cm) node (n3) {};
  \draw (0.7,0) ++(225:0.35cm) arc (225:315:0.35cm) node (n4) {}
        (0.7,0) ++(315:0.35cm) arc (315:360:0.35cm) node (n5) {}
        (n2.base)--+(145:0.5cm)
        (n3.base)--+(215:0.5cm)
        (n5.base)--+(0:0.5cm);
  \draw[name path=a] (n1.base)--(n3.base);
  \draw[white, name path=b] (n2.base)--(n4.base);
  \path[name intersections={of=a and b,by=i}];
  \node[fill=white, inner sep=1pt, rotate=45] at (i) {};
  \draw (n2.base)--(n4.base);
}
-\frac{7}{36}
\tikz[baseline=(vert_cent.base)]{
  \node (vert_cent) {\hspace{-13pt}$\phantom{-}$};
  \draw (0.7,0) ++(0:0.35cm) arc (0:30:0.35cm) node (n1) {}
        (0.7,0) ++(30:0.35cm) arc (30:105:0.35cm) node (n2) {}
        (0.7,0) ++(105:0.35cm) arc (105:135:0.35cm) node (n3) {}
        (0.7,0) ++(135:0.35cm) arc (135:225:0.35cm) node (n4) {}
        (0.7,0) ++(225:0.35cm) arc (225:360:0.35cm) node (n5) {}
        (n1.base) to[out=200,in=300] (n2.base)
        (n3.base)--+(145:0.5cm)
        (n4.base)--+(215:0.5cm)
        (n5.base)--+(0:0.5cm);
}
+\frac12
\tikz[baseline=(vert_cent.base)]{
  \node (vert_cent) {\hspace{-13pt}$\phantom{-}$};
  \draw (0.7,0) ++(0:0.35cm) arc (0:90:0.35cm) node (n1) {}
        (0.7,0) ++(90:0.35cm) arc (90:135:0.35cm) node (n2) {}
        (0.7,0) ++(135:0.35cm) arc (135:225:0.35cm) node (n3) {}
        (0.7,0) ++(225:0.35cm) arc (225:270:0.35cm) node (n4) {}
        (0.7,0) ++(270:0.35cm) arc (270:360:0.35cm) node (n5) {}
        (n1.base)--(n4.base)
        (n2.base)--+(145:0.5cm)
        (n3.base)--+(215:0.5cm)
        (n5.base)--+(0:0.5cm);
}
-\frac19
\tikz[baseline=(vert_cent.base)]{
  \node (vert_cent) {\hspace{-13pt}$\phantom{-}$};
  \draw (0,0.5)--(0.5,0)
        (0,-0.5)--(0.5,0)
        (0.5,0)--(0.8,0)
        (1.2,0) ++(0:0.4cm) arc (0:360:0.4cm and 0.3cm)
        (1.2,0.3)--(1.2,-0.3)
        (1.6,0)--(1.9,0);
}
+\frac{11}{216}
\tikz[baseline=(vert_cent.base)]{
  \node (vert_cent) {\hspace{-13pt}$\phantom{-}$};
  \draw (0,0.5)--(0.5,0)
        (0,-0.5)--(0.5,0)
        (0.5,0)--(0.8,0)
        (1.2,0) ++(0:0.4cm and 0.3cm) arc (0:50:0.4cm and 0.3cm) node (n1)
        {}
        (1.2,0) ++(50:0.4cm and 0.3cm) arc (50:130:0.4cm and 0.3cm) node
        (n2) {}
        (1.2,0) ++(130:0.4cm and 0.3cm) arc (130:360:0.4cm and 0.3cm)
        (n1.base) to[out=215,in=325] (n2.base)
        (1.6,0)--(1.9,0);
}
\right)\,.}[betatwo]
The first contribution to \betatwo is non-planar. For the seemingly
asymmetric vertex corrections in \betatwo (second and third term) a
symmetrization is understood; for example,
\eqn{
\tikz[baseline=(vert_cent.base)]{
  \node (vert_cent) {\hspace{-13pt}$\phantom{-}$};
  \draw (0.7,0) ++(0:0.35cm) arc (0:90:0.35cm) node (n1) {}
        (0.7,0) ++(90:0.35cm) arc (90:135:0.35cm) node (n2) {}
        (0.7,0) ++(135:0.35cm) arc (135:225:0.35cm) node (n3) {}
        (0.7,0) ++(225:0.35cm) arc (225:270:0.35cm) node (n4) {}
        (0.7,0) ++(270:0.35cm) arc (270:360:0.35cm) node (n5) {}
        (n1.base)--(n4.base)
        (n2.base)--+(145:0.5cm)
        (n3.base)--+(215:0.5cm)
        (n5.base)--+(0:0.5cm);
}\quad
\text{represents}
\quad
\tikz[baseline=(vert_cent.base)]{
  \node (vert_cent) {\hspace{-13pt}$\phantom{-}$};
  \draw (0.7,0) ++(0:0.35cm) arc (0:90:0.35cm) node (n1) {}
        (0.7,0) ++(90:0.35cm) arc (90:135:0.35cm) node (n2) {}
        (0.7,0) ++(135:0.35cm) arc (135:180:0.35cm) node (n3) {}
        (0.7,0) ++(180:0.35cm) arc (180:225:0.35cm) node (n4) {}
        (0.7,0) ++(225:0.35cm) arc (225:360:0.35cm) node (n5) {}
        (n1.base) to[out=270,in=0] (n3.base)
        (n2.base)--+(145:0.5cm)
        (n4.base)--+(215:0.5cm)
        (n5.base)--+(0:0.5cm);
}
+
\tikz[baseline=(vert_cent.base)]{
  \node (vert_cent) {\hspace{-13pt}$\phantom{-}$};
  \draw (0.7,0) ++(0:0.35cm) arc (0:90:0.35cm) node (n1) {}
        (0.7,0) ++(90:0.35cm) arc (90:135:0.35cm) node (n2) {}
        (0.7,0) ++(135:0.35cm) arc (135:225:0.35cm) node (n3) {}
        (0.7,0) ++(225:0.35cm) arc (225:270:0.35cm) node (n4) {}
        (0.7,0) ++(270:0.35cm) arc (270:360:0.35cm) node (n5) {}
        (n1.base)--(n4.base)
        (n2.base)--+(145:0.5cm)
        (n3.base)--+(215:0.5cm)
        (n5.base)--+(0:0.5cm);
}
+
\tikz[baseline=(vert_cent.base)]{
  \node (vert_cent) {\hspace{-13pt}$\phantom{-}$};
  \draw (0.7,0) ++(0:0.35cm) arc (0:135:0.35cm) node (n1) {}
        (0.7,0) ++(135:0.35cm) arc (135:180:0.35cm) node (n2) {}
        (0.7,0) ++(180:0.35cm) arc (180:225:0.35cm) node (n3) {}
        (0.7,0) ++(225:0.35cm) arc (225:270:0.35cm) node (n4) {}
        (0.7,0) ++(270:0.35cm) arc (270:360:0.35cm) node (n5) {}
        (n2.base) to[out=0,in=90] (n4.base)
        (n1.base)--+(145:0.5cm)
        (n3.base)--+(215:0.5cm)
        (n5.base)--+(0:0.5cm);
}\,.
}[]
In the single-field case \betatwo reproduces the result
of~\cite{Macfarlane:1974vp} (see also~\cite{Kodaira:1985vr, Kodaira:1985pg,
Jack:1985wf}\foot{There is a typo in the relevant equation in
\cite{Jack:1985wf}.}), which, just like $\beta^{(1)}$, is also negative.

The results presented here for the anomalous dimension and the beta
function to two loops are found to agree with the results of
\cite{Amit:1976pz, deAlcantaraBonfim:1980pe}, and can also be fully
extracted from \cite{deAlcantaraBonfim:1981sy}.

\newsec{The metric in coupling space and the
\texorpdfstring{$a$}{a}-anomaly}[metricSec]
In section~\ref{weylCCsSec}, and in particular in equation~\finalCC, it was
made apparent that there is an important piece of the $\beta_{\lambda}
\cdot \mathscr{R}$ terms, called $\cH^1_{IJ}$ in this paper and
in~\cite{Grinstein:2013cka},\foot{$\cH^1_{IJ}$ was called $\chi_{IJ}$ in
the two- and four-dimensional cases of~\cite{Osborn:1991gm} and
in the six-dimensional case of~\cite{Grinstein:2014xba}.} that manifests itself as the
coefficient of contact terms of certain correlation functions of the
operators of the theory in flat spacetime and spacetime-independent
$g_{ijk} (x)$ and $m_{ij} (x)$. This metric is important because it
controls the behavior of $a$ (or really, $\tilde{a}$) along the
renormalization group flow; given the outstanding importance given to $a$
(or its analogs) in two and four dimensions for its central role in
characterizing quantum field theories there, its behavior in six dimensions
gives insight into the universal features of quantum field theories in any
dimension, possibly beyond the conventional Lagrangian description so
ubiquitous in our understanding today.

In~\cite{Grinstein:2014xba} a perturbative computation of the theory
defined in the Lagrangian formalism by~\lagrangian yielded a surprising
result for the value of $\cH^1_{IJ}$ at the two loop level. One of the main
purposes of this work is to give the details of that computation, along
with other interesting results from the computation of the effective
action.

In order to compute $\cH^1_{IJ}$, we must first identify the corresponding
piece in $\lambda \cdot \mathscr{R}$ so that we may use \RGEct to obtain
the quantity of interest. There is no candidate in the one loop
computation, $\lambda^{(1)} \cdot \mathscr{R}$. Thus, we must look for a
two loop contribution in $\lambda^{(2)} \cdot \mathscr{R}$, where we do
indeed find a candidate. We see from \lambdaTwo that the relevant piece is
\eqn{\lambda^{(2)}\cdot\mathscr{R}\supset
-\frac{1}{\epsilon}\frac{1}{(64\pi^3)^2}\frac{1}{12\lsp 960}
H_1^{\mu\nu}\lsp\partial_\mu g_{ijk}\,\partial_\nu g_{ijk}\,.
}[AnomMetricAndW]
From \AnomMetricAndW and \TraceAnomDelta we can immediately match to the
term $\tfrac12\cH_{IJ}^1\lsp\partial_\mu g^I\lsp\partial_\nu g^J
H_1^{\mu\nu}$ in $\beta_\lambda\cdot\mathscr{R}$ and extract
\eqn{\cH^{1\lsp(2)}_{IJ}=-\frac{1}{(64\pi^3)^2}\frac{1}{3240}\delta_{IJ}\,,
}[ResMetric]
where, as section \methodOfCalc, we use notation of~\cite{Grinstein:2013cka} and denote $I=(ijk)$.
Furthermore, performing a Weyl variation of \AnomMetricAndW we find
\eqn{\delta_\sigma(-\lambda^{(2)}\cdot\mathscr{R})\supset
\frac{1}{\epsilon}\frac{1}{(64\pi^3)^2}\frac{1}{6480}
H_1^{\mu\nu}\hat{\beta}_{ijk}\,\partial_\nu g_{ijk}\,\partial_\mu\sigma\,,
}[]
and so
\eqn{\cH_{I}^{1\lsp(2)}=-\frac{1}{(64\pi^3)^2}\frac{1}{12\lsp960}g_I\,,
}[ResW]
as also seen in~(\ref{eq:Zmu}). The result \ResMetric is unambiguous and
scheme-independent.  As we observe, the leading, two-loop contribution to
the metric is negative, and so the consistency condition \aCC and its
consequence \finalCC cannot possibly lead to a strong $a$-theorem for
$\tilde{a}$.

Now, our theory has only the Gaussian fixed point in perturbation theory.
Non-perturbatively there may be a non-trivial fixed point, but our results
\ResMetric and \ResW cannot be used beyond perturbation theory.
Nevertheless, as long as the flow of our theory can be described
perturbatively, the quantity $\tilde{a}$ is monotonically increasing.

Another use of the consistency conditions is the evaluation of some
quantities at higher loop orders. Regarding $\tilde{a}$, for example, we
can use \aCC with the results \betaone, \ResMetric, and \ResW to obtain the
three-loop contribution to $\tilde{a}$,
\eqn{\tilde{a}^{(3)}=\frac{1}{(64\pi^3)^3}\frac{1}{77\hspace{1pt}760}
\left(
\tikz[baseline=(vert_cent.base)]{
  \node (vert_cent) {\hspace{-13pt}$\phantom{-}$};
  \draw (0,0) circle [radius=0.35cm];
  \draw (0,0)--(120:0.35cm)
        (0,0)--(240:0.35cm)
        (0,0)--(0:0.35cm);
}-
\frac14\,
\tikz[baseline=(vert_cent.base)]{
  \node (vert_cent) {\hspace{-13pt}$\phantom{-}$};
  \draw (0,0) circle [x radius=0.45cm, y radius=0.35cm];
  \draw (-0.45,0)--(-0.2,0)
        (0.45,0)--(0.2,0);
  \draw (0,0) circle [x radius=0.2cm, y radius=0.15cm];
}
\right).}[atildethree]

Furthermore, from the consistency conditions (see \cite{Grinstein:2013cka}
for the meaning of the various terms)
\twoseqn{b_1&=\tfrac16(\mathcal{F}_I-\tfrac12\partial_Ib_{14}-\tfrac12\mathcal{I}_I^7)
\beta^I,}[DHFour]{b_3&=(\mathcal{F}_I+\partial_Ib_{13}
-\partial_Ib_{14}+\mathcal{I}_I^6-\mathcal{I}_I^7)\beta^I,}[DHThree][DD]
it is clear that at two loops $b_1^{(2)}=b_3^{(2)}=0$. This, in conjunction
with \aCC, \ResMetric and \ResW, implies that $a^{(2)}=0$. These results
have been verified by our explicit computations~(\ref{eq:FTwo}). Now, at two loops we
can use~(\ref{eq:FTwo}) and~(\ref{eq:Zmu}) to obtain
\eqn{\mathcal{F}_{I}^{(2)}=\frac{1}{(64\pi^3)^2}\frac{1}{1080}g_{I}\,,
\qquad b_{13}^{(2)}=b_{14}^{(2)}=0\,,\qquad
\mathcal{I}_{I}^{6\lsp(2)}=\mathcal{I}_{I}^{7\lsp(2)}=0\,,}[]
and so using \DD we can compute
\eqn{b_1^{(3)}=\tfrac16 b_3^{(3)}=
-\frac{1}{(64\pi^3)^3}\frac{1}{6480}\left(
\tikz[baseline=(vert_cent.base)]{
  \node (vert_cent) {\hspace{-13pt}$\phantom{-}$};
  \draw (0,0) circle [radius=0.35cm];
  \draw (0,0)--(120:0.35cm)
        (0,0)--(240:0.35cm)
        (0,0)--(0:0.35cm);
}-
\frac14\,
\tikz[baseline=(vert_cent.base)]{
  \node (vert_cent) {\hspace{-13pt}$\phantom{-}$};
  \draw (0,0) circle [x radius=0.45cm, y radius=0.35cm];
  \draw (-0.45,0)--(-0.2,0)
        (0.45,0)--(0.2,0);
  \draw (0,0) circle [x radius=0.2cm, y radius=0.15cm];
}
\right).}[]
With these results and using \aCC with \betaone, \ResMetric, and \ResW we
find that the three-loop contribution to $a$ is
\eqn{a^{(3)}=\frac{1}{(64\pi^3)^3}\frac{1}{64\hspace{1pt}800}\left(
\tikz[baseline=(vert_cent.base)]{
  \node (vert_cent) {\hspace{-13pt}$\phantom{-}$};
  \draw (0,0) circle [radius=0.35cm];
  \draw (0,0)--(120:0.35cm)
        (0,0)--(240:0.35cm)
        (0,0)--(0:0.35cm);
}-
\frac14\,
\tikz[baseline=(vert_cent.base)]{
  \node (vert_cent) {\hspace{-13pt}$\phantom{-}$};
  \draw (0,0) circle [x radius=0.45cm, y radius=0.35cm];
  \draw (-0.45,0)--(-0.2,0)
        (0.45,0)--(0.2,0);
  \draw (0,0) circle [x radius=0.2cm, y radius=0.15cm];
}
\right).}[athree]
This shows that, just like $\tilde{a}$, $a$ increases in the flow out of
the trivial UV fixed point in our theory.

There is one comment to be made about the value of the result in
\atildethree.  It is a scheme-dependent quantity, in the sense that it is
only defined modulo terms that are ``exact'' in the cohomology generated by
the Weyl transformations $\DeltaW_{\sigma} - \Delta^{\beta}_{\sigma}$,
i.e.\ up to local additions to the original action whose variations shift
quantities in \finalCC. However, as shown in~\cite{Grinstein:2013cka} in
analogy with~\cite{Osborn:1991gm} these shifts are of the form $\delta
\tilde{a} = z_{IJ} \beta^I \beta^J$ for $z_{IJ}$ an arbitrary regular
symmetric function of the couplings. Hence, at lowest order, and using
\betaone, we have $\delta \tilde{a} \sim \mathcal{O} (g^6)$ which cannot
possibly upset the conclusions of this section in perturbation theory.
Moreover, equation \finalCC is of course unchanged by such shifts and in
this sense is an invariant of the associated Weyl cohomology.

%%fakesection Acknowledgments
\ack{We are grateful to Hugh Osborn for his careful reading of the
manuscript and for his useful comments and insights. We have relied heavily
on \emph{Mathematica} and the package
\href{http://www.xact.es/}{\texttt{xAct}}. The research of BG is supported
in part by the Department of Energy under grant DE-SC0009919. The research
of AS is supported in part by the National Science Foundation under Grant
No.~1350180. The research of DS is supported by a grant from the European
Research Council under the European Union's Seventh Framework Programme (FP
2007-2013) ERC Grant Agreements No.~279972 ``NPFlavour.'' The research of
MZ is supported in part by the National Science Foundation of China under
Grants No.\ 11475258 and No.\ 11205242.}

%%fakesection Appendices
\begin{appendices}
\newsec{Conventions and basis tensors}[appConv]
In this work we define the Riemann tensor via
\eqn{[\nabla_{\mu},\nabla_{\nu}] A^{\rho} =
R^\rho{\!}_{\sigma\mu\nu} A^{\sigma} \,,}[mswConvention]
and the Ricci tensor and Ricci scalar as
$R_{\mu\nu}=R^{\rho}{\!}_{\mu\rho\nu}$ and $R=\gamma^{\mu\nu}R_{\mu\nu}$.
We also commonly use the Weyl tensor defined in $d\geq3$ by
\eqn{W_{\mu\nu\rho\sigma}=R_{\mu\nu\rho\sigma}+
\tfrac{2}{d-2}(\gamma_{\mu[\sigma}R_{\rho]\nu}+
\gamma_{\nu[\rho}R_{\sigma]\mu})+
\tfrac{2}{(d-1)(d-2)}\gamma_{\mu[\rho}\gamma_{\sigma]\nu}R\,.}[]
At mass dimension four we use the tensors
\eqn{\begin{gathered}
E_4=\tfrac{2}{(d-2)(d-3)}(R^{\mu\nu\rho\sigma}R_{\mu\nu\rho\sigma}
  -4R^{\mu\nu}R_{\mu\nu}+R^2)\,,\\
F=W^{\mu\nu\rho\sigma}W_{\mu\nu\rho\sigma}\,,\qquad
\tfrac{1}{(d-1)^2}R^2\,,\qquad
\tfrac{1}{d-1}\nabla^2 R\,,\\
H_{1\llsp\mu\nu}=\tfrac{(d-2)(d-3)}{2}E_4\gamma_{\mu\nu}-4(d-1)H_{2\mu\nu}
  +8H_{3\mu\nu}+8H_{4\mu\nu}
  -4R^{\rho\sigma\tau}{\!}_{\mu}R_{\rho\sigma\tau\nu}\,,\\
H_{2\llsp\mu\nu}=\tfrac{1}{d-1}RR_{\mu\nu}\,,\qquad
H_{3\llsp\mu\nu}=R_{\mu}{\!}^{\rho}R_{\rho\nu}\,,\qquad
H_{4\llsp\mu\nu}=R^{\rho\sigma}R_{\rho\mu\sigma\nu}\,,\\
H_{5\llsp\mu\nu}=\nabla^2 R_{\mu\nu}\,,\qquad
H_{6\llsp\mu\nu}=\tfrac{1}{d-1}\nabla_\mu\partial_\nu R\,.
\end{gathered}}[BFour]
A complete basis of scalar dimension-six curvature terms consists of \cite{Bonora:1985cq}
\eqn{\begin{gathered}
K_1=R^3\,,\qquad
K_2=RR^{\mu\nu}R_{\mu\nu}\,,\qquad
K_3=RR^{\mu\nu\rho\sigma}R_{\mu\nu\rho\sigma}\,,\qquad
K_4=R^{\mu\nu}R_{\nu\rho}R^\rho_{\hphantom{\rho}\!\mu}\,,\\
K_5=R^{\mu\nu}R^{\rho\sigma}R_{\mu\rho\sigma\nu}\,,\qquad
K_6=R^{\mu\nu}R_{\mu\rho\sigma\tau}
  R_{\nu}^{\smash{\hphantom{\nu}\rho\sigma\tau}}\,,\qquad
K_7=R^{\mu\nu\rho\sigma}R_{\rho\sigma\tau\omega}
R^{\tau\omega}{\!}_{\mu\nu}\,,\\
K_8=R^{\mu\nu\rho\sigma}R_{\tau\nu\rho\omega}
R_{\mu}{\!}^{\tau\omega}{\!}_\sigma\,,\qquad
K_9=R\,\nabla^2 R\,,\qquad
K_{10}=R^{\mu\nu}\,\nabla^2 R_{\mu\nu}\,,\qquad
K_{11}=R^{\mu\nu\rho\sigma}\,\nabla^2 R_{\mu\nu\rho\sigma}\,,\\
K_{12}=R^{\mu\nu}\nabla_\mu\partial_\nu R\,,\qquad
K_{13}=\nabla^\mu R^{\nu\rho}\lsp\nabla_{\mu}R_{\nu\rho}\,,\qquad
K_{14}=\nabla^\mu R^{\nu\rho}\lsp\nabla_{\nu}R_{\mu\rho}\,,\\
K_{15}=\nabla^\mu R^{\nu\rho\sigma\tau} \lsp
\nabla_\mu R_{\nu\rho\sigma\tau}\,,\qquad
K_{16}=\nabla^2 R^2\,,\qquad
K_{17}=(\nabla^2)^2 R\,.
\end{gathered}}

In $d=6$ a convenient basis is given by
\eqna{I_1&=\tfrac{19}{800}\llsp K_1 - \tfrac{57}{160}\llsp K_2 + \tfrac{3}{40}\llsp K_3 +
\tfrac{7}{16}\llsp K_4 - \tfrac{9}{8}\llsp K_5 - \tfrac{3}{4}\llsp K_6 +
\llsp K_8\,,\\
I_2&=\tfrac{9}{200}\llsp K_1 - \tfrac{27}{40}\llsp K_2 + \tfrac{3}{10}\llsp K_3 +
\tfrac{5}{4}\llsp K_4 - \tfrac{3}{2}\llsp K_5 - 3\llsp K_6 + \llsp K_7\,,\\
I_3&=-\tfrac{11}{50}\llsp K_1 + \tfrac{27}{10}\llsp K_2 - \tfrac{6}{5}\llsp K_3 - \llsp K_4 + 6\llsp K_5+
2\llsp K_7 - 8\llsp K_8\\&\hspace{3cm} + \tfrac{3}{5}\llsp K_9 - 6\llsp K_{10} + 6\llsp K_{11} + 3\llsp K_{13}
- 6\llsp K_{14} + 3\llsp K_{15}\,,\\
E_6&=\llsp K_1 - 12\llsp K_2 + 3\llsp K_3 + 16\llsp K_4 - 24\llsp K_5 - 24\llsp K_6 + 4\llsp K_7 +
8\llsp K_8\,,\\
J_1&=6\llsp K_6 - 3\llsp K_7 + 12\llsp K_8 + \llsp K_{10} - 7\llsp K_{11} - 11\llsp K_{13} + 12\llsp K_{14} -
  4\llsp K_{15}\,,\\
J_2&=-\tfrac15 \llsp K_9 + \llsp K_{10} + \tfrac25 \llsp K_{12} + \llsp K_{13}\,,\qquad
J_3=\llsp K_4 + \llsp K_5 - \tfrac{3}{20}\llsp K_9 + \tfrac45 \llsp K_{12}
  + \llsp K_{14}\,,\\
J_4&=-\tfrac15 \llsp K_9 + \llsp K_{11} + \tfrac25 \llsp K_{12} + \llsp K_{15}\,,\qquad
J_5=\llsp K_{16}\,,\qquad
J_6=\llsp K_{17}\,,\\
L_{1}&=-\tfrac{1}{30}\llsp K_1+\tfrac14 \llsp K_2-\llsp K_6\,,\qquad
L_{2}=-\tfrac{1}{100}\llsp K_1+\tfrac{1}{20}\llsp K_2\,,\\
L_{3}&=-\tfrac{37}{6000}\llsp K_1+\tfrac{7}{150}\llsp K_2 -\tfrac{1}{75}\llsp K_3
+\tfrac{1}{10}\llsp K_5+\tfrac{1}{15}\llsp K_6\,,\qquad
L_{4}=-\tfrac{1}{150}\llsp K_1+\tfrac{1}{20}\llsp K_3\,,\\
L_{5}&=\tfrac{1}{30}\llsp K_1\,,\qquad
L_{6}=-\tfrac{1}{300}\llsp K_1+\tfrac{1}{20}\llsp K_9\,,\qquad
L_{7}=\llsp K_{15}\,,}[]
where the first three transform covariantly under Weyl variations, and
$E_6$ is the Euler term in $d=6$.  The $J$'s are trivial anomalies in a
six-dimensional CFT defined in curved space, and the first six $L$'s are
constructed based on the relation $\delta_\sigma\int \dsix x\sqrt{\gamma}\,
L_{1,\ldots,6}=\int \dsix x\sqrt{\gamma}\,\sigma J_{1,\ldots,6}$.

In this paper we use the above basis for dimension-six curvature scalars,
but, although it is not necessary, we define $I_{1,2,3}$ in general $d$,
because of our use of dimensional regularization. More specifically, we
define
\begin{align}
  I_1&=W^{\mu\nu\rho\sigma}W_{\tau\nu\rho\omega}
W_\mu{\!}^{\tau\omega}{\!}_\sigma\nonumber\\
  &=\frac{\dtwo+d-4}{(d-1)^2(d-2)^3}\lsp K_1
  -\frac{3\lsp(\dtwo+d-4)}{(d-1)(d-2)^3}\lsp K_2
  +\frac{3}{2\lsp (d-1)(d-2)}\lsp K_3 +\frac{2\lsp(3\lsp d-4)}{(d-2)^3}
  \lsp K_4\nonumber\\
  &\hspace{9cm}-\frac{3\lsp d}{(d-2)^2}\lsp K_5 -\frac{3}{d-2}\lsp K_6
  +K_8\,,\displaybreak[0]\nonumber\\
I_2&=W^{\smash{\mu\nu\rho\sigma}}W_{\rho\sigma\tau\omega}
  W^{\tau\omega}{\!}_{\mu\nu}\nonumber\\
  &=\frac{8\lsp(2\lsp d-3)}{(d-1)^2(d-2)^3}\lsp K_1
  -\frac{24\lsp(2\lsp d-3)}{(d-1)(d-2)^3}\lsp K_2
  +\frac{6}{(d-1)(d-2)}\lsp K_3
  +\frac{16\lsp(d-1)}{(d-2)^3}\lsp K_4\nonumber\\
  &\hspace{9cm}-\frac{24}{(d-2)^2}\lsp K_5 -\frac{12}{d-2}\lsp
  K_6+K_7\,,\displaybreak[0]\nonumber\\
I_3&=W^{\mu\nu\rho\sigma}\left(\delta_{\mu}
{\!}^{\tau}\,\nabla^2 + \frac{16}{d-2}R_{\mu}
{\!}^{\tau} -\frac{4\lsp d}{(d-1)(d-2)}\delta_{\mu}
{\!}^{\tau}R\right)
  W_{\tau\nu\rho\sigma}\nonumber\\
  &\hspace{6cm}+8\lsp \nabla^\mu\nabla^\nu
  (W_{\mu}{\!}^{\rho\sigma\tau}W_{\nu\rho\sigma\tau})-\frac12\lsp\nabla^2(W^{\mu\nu\rho\sigma}W_{\mu\nu\rho\sigma})\nonumber\\
  &=-\frac{2\lsp(\dtwo+d+2)}{(d-1)^2(d-2)^2}\lsp K_1
  +\frac{2\lsp(\dtwo+13\lsp d-6)}{(d-1)(d-2)^2}\lsp K_2
  -\frac{2\lsp(d-3)}{d-1}\lsp K_3
  +\frac{4\lsp (d-10)}{(d-2)^2}\lsp K_4\nonumber\\
  &\quad-\frac{6\lsp (d-10)}{d-2}\lsp K_5
  -\frac12\lsp (d-10)\lsp K_7
  +2\lsp(d-10)K_8
  -\frac{(d-3)(d-10)}{(d-1)(d-2)}\lsp K_9\nonumber\\
  &\quad-\frac{8\lsp(d-3)}{d-2}\lsp K_{10}
  +2\lsp (d-3)\lsp K_{11}
  -\frac{(d-3)(3\lsp d-22)}{d-2}\lsp K_{13}
  +\frac{2\lsp(d-3)(d-10)}{d-2}\lsp K_{14}\nonumber\\
  &\quad+\frac14\lsp(d-2)(d-3)K_{15}
  +\frac{(d-3)(d-6)}{2\lsp(d-1)(d-2)}K_{16}\,.
\end{align}
These satisfy $\delta_\sigma I_{1,2,3}=6\lsp\sigma I_{1,2,3}$ for any
$d$ for which they can be defined.

\newsec{Coincident limits}[appLimits]
Here we collect the coincident limits $x' \rightarrow x$ of the
Seeley--DeWitt coefficients $a_{n,\lsp ij}(x,x')$ of \hkansatz and the various
functions (i.e.\ $\sigma (x,x')$ and $\dVM (x,x')$) needed therein to solve
the recursion relation \recursionrelation. Most of these results can be
found in \cite{Jack:1985wf,
Kodaira:1985pg}\foot{Reference~\cite{Kodaira:1985pg} uses the opposite
curvature convention as the one used in this work, \mswConvention.  This
must be taken into account when comparing expressions, as odd powers of
curvature will have an extra relative minus sign.}, and~\cite{Jack:1983sk},
though in these works only the single coupling case was considered.

The fundamental quantities of interest on a curved background are the
geodetic interval $\sigma (x,x')$, whose ``equation of motion''
is~\cite{DeWitt:1965jb}
\eqn{\tfrac{1}{2}\lsp\partial^{\mu} \sigma \lsp \partial_{\mu} \sigma =
\sigma\,,}[fundamentalEqnA]
and the van Vleck--Morette determinant $\Delta_{\text{VM}}(x,x')$ , which
describes the rate at which geodesics coming from a point separate, follows
from a corollary of the above equation for $\sigma (x,x')$:
\eqn{ \dVM \lsp\nabla^2 \sigma + 2 \lsp \partial^{\mu}\sigma \lsp
\partial_{\mu} \dVM = d \lsp \dVM\,,}[fundamentalEqnB]
with $d$ the spacetime dimension. Here we have abbreviated $\sigma (x,x')$
and $\dVM (x,x')$ as $\sigma$ and $\dVM$, respectively, and will continue
to do so throughout the rest of the appendix.

From these two equations we can construct the coincident limits of $\sigma$
and $\dVM$. We will denote the coincident limit of a quantity $X$ as $[ X
]$ for brevity.  Throughout the rest of this appendix, to make the
coincident limits as concise as possible, we use the semicolon notation for
the covariant derivatives, e.g. $\nabla_{\nu} \partial_{\mu} \sigma =
\sigma_{\llsp;\llsp\mu\nu}$.  Now, since $\sigma$ measures a distance, we
clearly have
\eqn{[\sigma] = 0 \,.}[]
Now using \fundamentalEqnA and differentiating as many times as needed, we
obtain the following limits:
\eqna{
	[ \sigma_{\llsp,\llsp\mu} ] &= \lsp 0 \,, \qquad [
	\sigma_{\llsp;\llsp\mu\nu} ]  = \gamma_{\mu\nu} \,, \qquad [
	\sigma_{\llsp;\llsp\mu\nu\rho} ] = 0\,,
	\\
	[ \sigma_{\llsp;\llsp\llsp\mu\nu\rho\sigma} ]  &=
	-\tfrac{1}{3} ( R_{\mu\rho\nu\sigma} +
	R_{\mu\sigma\nu\rho} )	\,,
	\\
	[ \sigma_{\llsp;\llsp\mu\nu\rho\sigma\tau} ] &=
	-\tfrac{1}{4}( R_{\mu\rho\nu\sigma\llsp;\llsp\tau} +
	R_{\mu\sigma\nu\rho\llsp;\llsp\tau} + \tau \leftrightarrow
	\sigma + \tau \leftrightarrow \rho ) \,,
	\\
    [\sigma_{\llsp;\llsp\rho}{}^\rho{\!}_\sigma
    {}^\sigma{\!}_{\mu \nu}]&=
    \tfrac{4}{45}\lsp R_{\mu \rho}R^{\rho}{\!}_{\nu} +
	\tfrac{8}{45}\lsp R^{\rho \sigma}R_{\rho\mu\sigma\nu} -
    \tfrac{4}{15}\lsp R_{\mu}{\!}^{\rho\sigma\tau}
    R_{\nu\rho\sigma\tau} - \tfrac{2}{5}\lsp
    R_{\mu\nu\llsp;\llsp\rho}{}^{\rho} -
    \tfrac{6}{5}\lsp R_{\lsp;\llsp\mu\nu}\,,
	\\
    [\sigma_{\llsp;\llsp\rho}{}^\rho{\!}_{\mu\nu\sigma}
    {\hspace{-0.5pt}}^\sigma]&=
    [\sigma_{\llsp;\llsp\rho}{}^\rho{\!}_\sigma
    {}^\sigma{\!}_{\mu \nu}]+ \tfrac{4}{3}
    ( R^{\rho\sigma}R_{\rho\mu\sigma\nu} -
    R_{\mu}{\!}^\rho R_{\rho\nu} ) \,,
	\\
    [\sigma_{\llsp;\llsp\mu\nu\rho}{}^{\rho}
    {\!}_\sigma{}^\sigma]&=
    [\sigma_{\llsp;\llsp\rho}{}^\rho{\!}_{\mu\nu\sigma}
    {\hspace{-0.5pt}}^\sigma] - \tfrac{8}{3} ( R^{\rho\sigma}
    R_{\rho\mu\sigma\nu} -
    R_{\mu}{\!}^\rho R_{\rho\nu} )
    -2\lsp R_{\mu\nu\llsp;\llsp\rho}{}^{\rho}
    + 2\lsp R_{\lsp;\llsp\mu\nu} \,,
    \\
    [\sigma_{\llsp;\llsp\mu}{}^\mu{\!}_\nu{}^\nu
    {\!}_\rho{}^\rho] &=\tfrac45\lsp(\tfrac13\lsp
    R^{\mu\nu}R_{\mu\nu} -\tfrac13\lsp
    R^{\mu\nu\rho\sigma} R_{\mu\nu\rho\sigma}-2\lsp
    R_{\lsp;\llsp\mu}{\!}^\mu)\,,
	\\
    [\sigma_{\llsp;\llsp\mu}{}^\mu{\!}_\nu{}^\nu
    {\!}_\rho{}^\rho{\!}_\sigma{}^\sigma] &=
    -\tfrac{152}{315}\llsp K_{4}-\tfrac{176}{315}\llsp
    K_5+\tfrac{296}{315}\llsp K_6-\tfrac{44}{63}\llsp K_7
    +\tfrac{80}{63}\llsp K_8+\tfrac{3}{7}\llsp K_9
    +\tfrac{148}{105}\llsp K_{10}\\
    &\quad-\tfrac{12}{7}\llsp K_{11}-\tfrac{32}{35}\llsp
    K_{12}+\tfrac{2}{7}\llsp K_{13}+\tfrac{4}{7}\llsp K_{14}
    -\tfrac{9}{7}\llsp K_{15}-\tfrac{3}{14}\llsp K_{16}
    -\tfrac{18}{7}\llsp K_{17}\,.
}[sigmaLimits]%
The coincident limits of derivatives on $\dVM$ follow from
\fundamentalEqnB:
\eqna{
[ \dVM ] &= 1 \,, \qquad [ \dVM{}_{\llsp,\llsp\mu}^{\phantom{\mu}}] = 0
	\,, \qquad [ \dVM{}_{\llsp;\llsp\mu\nu}^{\phantom{\mu}} ]
    = \tfrac{1}{6}\lsp R_{\mu\nu} \,,
	\\
	[ \dVM{}_{\llsp;\llsp\mu\nu\rho}^{\phantom{\mu}} ] &=
    \tfrac{1}{12} (R_{\mu\nu\llsp;\llsp\rho}
    + R_{\rho\mu\llsp;\llsp\nu} + R_{\nu\rho\llsp;\llsp\mu}
	) \,,
	\\
    [\dVM{}_{\llsp;\llsp\rho}^{\phantom{\mu}}{}^\rho
    {\!}_{\mu\nu}^{\phantom{\mu}}]
    &= \tfrac{1}{36}\llsp R R_{\mu\nu} - \tfrac{1}{15}\llsp
    R_{\mu}{\!}^\rho R_{\rho\nu} + \tfrac{1}{30} (
	R^{\rho\sigma} R_{\rho\mu\sigma\nu} +
    R_{\mu}{\!}^{\rho\sigma\tau}R_{\nu\rho\sigma\tau} )
    + \tfrac{3}{20}\llsp R_{\lsp;\llsp\mu\nu} +
    \tfrac{1}{20}\llsp R_{\mu\nu\llsp;\llsp\rho}{}^{\rho} \,,
	\\
	[\dVM{}_{\llsp;\llsp\mu\nu\rho}^{\phantom{\mu}}{}^\rho]&=
    [\dVM{}_{\llsp;\llsp\rho}^{\phantom{\mu}}{}^\rho{\!}
    _{\mu\nu}^{\phantom{\mu}}]
    + \tfrac{1}{3} (R_{\mu}{\!}^\rho R_{\rho\nu} -
	R^{\rho\sigma}R_{\rho\mu\sigma\nu} )\,,
	\\
    [\dVM{}_{\llsp;\llsp\mu}^{\phantom{\mu}}{}^\mu
    {\!}_\nu^{\phantom{\mu}}{}^\nu{\!}_\rho^{\phantom{\mu}}{}^\rho] &=
    \tfrac{1}{216}\llsp K_1-\tfrac{1}{60}\llsp K_2+\tfrac{1}{60}\llsp K_3
    +\tfrac{2}{189}\llsp K_4+\tfrac{2}{63}\llsp K_5-\tfrac{4}{63}\llsp K_6
    +\tfrac{11}{189}\llsp K_7-\tfrac{20}{189}\llsp K_8-\tfrac{2}{105}
    \llsp K_9\\
    &\quad-\tfrac{2}{21}\llsp K_{10}+\tfrac{1}{7}\llsp K_{11}
    +\tfrac{1}{7}\llsp K_{12}-\tfrac{1}{42}\llsp K_{13}
    -\tfrac{1}{21}\llsp K_{14}+\tfrac{3}{28}\llsp K_{15}
    +\tfrac{5}{84}\llsp K_{16}+\tfrac{3}{14}\llsp K_{17}\,.
}[dVMLimits]
We have only listed the limits that are needed to compute the coincident
limits of the Seeley--DeWitt coefficients $a_{0,\lsp ij}$ up to $a_{3,\lsp ij}$.
We also note that we have explicitly checked all limits with those found in
the aforementioned references and find agreement.

Another quantity of interest, which is indirectly related to our
calculations in this paper through the details laid out in
appendix~\ref{appProp}, is $Y
\equiv\Delta^{-1/2}_{\text{VM}}\dVM{}_{\llsp;\llsp\mu}^{\phantom{Z}}
{}^\mu$.  Knowledge of its coincident limits is necessary for the
computation of \twoLoopGraph. We find
\eqna{
[Y_{\llsp,\llsp\mu}]&=\tfrac16\llsp R_{\lsp,\llsp\mu}\,,\\
[Y_{\llsp;\llsp\mu\nu}]&= - \tfrac{1}{15}\llsp
R_{\mu}{\!}^\rho R_{\rho\nu} + \tfrac{1}{30} (
R^{\rho\sigma} R_{\rho\mu\sigma\nu} +
R_{\mu}{\!}^{\rho\sigma\tau}R_{\nu\rho\sigma\tau} )
+ \tfrac{3}{20}\llsp R_{\lsp;\llsp\mu\nu} +
\tfrac{1}{20}\llsp R_{\mu\nu\llsp;\llsp\rho}{}^{\rho} \,,\\
[Y_{\llsp;\llsp\mu}{}^\mu]&=-\tfrac{1}{30}(R^{\mu\nu}R_{\mu\nu}
-R^{\mu\nu\rho\sigma}R_{\mu\nu\rho\sigma})
+\tfrac15\llsp R_{\lsp;\llsp\mu}{}^\mu\,,\\
[Y_{\llsp;\llsp\nu}{}^\nu{\!}_\mu]&=-\tfrac{1}{15}(R^{\nu\rho}
R_{\nu\rho\llsp;\llsp\mu}-R^{\nu\rho\sigma\tau}
R_{\nu\rho\sigma\tau\llsp;\llsp\mu})+\tfrac15 \llsp
R_{\lsp;\llsp\nu}{}^{\nu}{\!}_\mu\,,\\
[Y_{\llsp;\llsp\mu\nu}{}^\nu]&=[Y_{\llsp;\llsp\nu}{}^\nu{\!}_\mu]
+\tfrac16\llsp R_\mu{\!}^\nu R_{\lsp,\llsp\nu}\,,\\
[Y_{\llsp;\llsp\mu}{}^\mu{\!}_\nu{}^\nu]&=\tfrac{52}{945}\llsp K_{4}
+\tfrac{17}{315}\llsp K_5-\tfrac{3}{35}\llsp K_6+\tfrac{11}{189}\llsp K_7-
\tfrac{20}{189}\llsp K_8-\tfrac{1}{126}\llsp K_9-\tfrac{9}{70}\llsp K_{10}
\\
&\quad+\tfrac{1}{7}\llsp K_{11}+\tfrac{3}{70}\llsp K_{12}-\tfrac{1}{42}
\llsp K_{13}-\tfrac{1}{21}\llsp K_{14}+\tfrac{3}{28}\llsp K_{15}
+\tfrac{1}{252}\llsp K_{16}+\tfrac{3}{14}\llsp K_{17}\,.
}[YLimits]

Now we may proceed to the quantities that are directly related to the
central computations of this paper, the Seeley--DeWitt coefficients
$a_{n,\lsp ij} (x,x')$ that characterize the propagator's response to the
curved background with metric $\gamma_{\mu\nu} (x)$. Restating its
fundamental and defining condition, \recursionrelation, we have
\eqn{n\hspace{0.5pt} a_{n,\lsp ij}+\partial_{\mu}\sigma\,
\partial^{\mu}a_{n,\lsp ij} =
-\Delta_{\text{VM}}^{-1/2} M_{ik} \big(\Delta_{\text{VM}}^{1/2}
a_{n-1,kj}\big)\,,}[recursionRelationApp]
with initial conditions
\eqn{\partial_{\mu}\sigma\,\partial^{\mu}a_{0,\lsp ij}=0 \qquad \text{and}
\qquad \left[ a_{0,\lsp ij} \right] = \delta_{ij}\,.}[]
The limits of these coefficients depend on the elliptic differential
operator of the form $M_{ij} = -\delta_{ij} \nabla^2 + X_{ij}$, with
$X_{ij} = \left. \frac{\partial^2
V(\phi)}{\partial\phi^i\,\partial\phi^j}\right|_{\phi=\phi_b}$. In the case of
the Lagrangian \LtildeWithM, we have
\eqn{ X_{ij} = m_{ij} + g_{ijk}\llsp \phi_k +
\xi_{ij} R\,.}[]
(Unless explicitly stated we will take $\phi$ to represent the background
field $\phi_b$ for the sake of compressed notation.) Note that $X_{ij}$ is
symmetric, $X_{ij} = X_{ji}$.  Then,
\eqna{
	[a_{1,\lsp ij}] & = \tfrac{1}{6}\llsp R\lsp \delta_{ij}  - X_{ij}\,,
	\\
	[\partial_{\mu} a_{1,\lsp ij}] & = \tfrac{1}{2} \partial_{\mu}
	\big( \tfrac{1}{6}\llsp R\lsp\delta_{ij} - X_{ij} \big)\,,
	\\
	[\nabla_{\mu} \partial_{\nu} a_{1,\lsp ij}] & =
	\tfrac{1}{45} \big( \tfrac{1}{2}(
    R^{\rho\sigma}R_{\rho\mu\sigma\nu} + R_{\mu}{\!}^{\rho\sigma\tau}
    R_{\nu\rho\sigma\tau})
    - R_{\mu}{\!}^{\rho} R_{\rho\nu}
    + \tfrac{3}{4}\lsp \nabla^2 R_{\mu\nu} +
	\tfrac{9}{4}\lsp \nabla_{\mu}\partial_{\nu} R \big) \delta_{ij}
	- \tfrac{1}{3}\lsp \nabla_{\mu} \partial_{\nu} X_{ij}\,,\\
    [\partial_{\mu}\nabla^2a_{1,\lsp ij}]&=
    \tfrac{1}{60}
    (R^{\nu\rho\sigma\tau}\lsp\nabla_\mu R_{\nu\rho\sigma\tau}
    -R^{\nu\rho}\lsp\nabla_\mu R_{\nu\rho}
    -\tfrac53\llsp R_\mu{\!}^{\nu}\lsp\partial_\nu R
    +3\lsp\partial_\mu\nabla^2R)\delta_{ij}
    -\tfrac14\llsp\partial_\mu\nabla^2 X_{ij}\,,\\
    [\nabla^2\partial_{\mu}a_{1,\lsp ij}]&=[\partial_{\mu}\nabla^2a_{1,\lsp ij}]
    +\tfrac{1}{2}\llsp R_{\mu}{\!}^\nu\lsp\partial_\nu
    (\tfrac16\llsp R\lsp\delta_{ij}- X_{ij})\,,\\
    [(\nabla^2)^2 a_{1,\lsp ij}]&=(\tfrac{16}{945} \llsp K_{4}
    +\tfrac{13}{945}\llsp K_{5} -\tfrac{19}{945} \llsp K_{6}
    +\tfrac{11}{945} \llsp K_{7}-\tfrac{4}{189} \llsp K_{8}
    +\tfrac{19}{1260} \llsp K_{9}-\tfrac{19}{630} \llsp K_{10}
    +\tfrac{1}{35} \llsp K_{11}\\
    &\quad -\tfrac{1}{210} \llsp K_{12} -\tfrac{1}{210} \llsp K_{13}
    -\tfrac{1}{105}\llsp K_{14} +\tfrac{3}{140} \llsp K_{15}
    -\tfrac{19}{2520}\llsp K_{16}+\tfrac{3}{70} \llsp K_{17})\delta_{ij}\\
    &\quad+\tfrac{4}{45}\llsp R^{\mu\nu}\lsp\nabla_\mu\partial_\nu X_{ij}
    +\tfrac{1}{10}\llsp\partial^\mu R\,\partial_\mu X_{ij}
    -\tfrac15\llsp(\nabla^2)^2X_{ij}\,.
}[aILimits]
for the relevant limits of $a_{1,\lsp ij}$. For $a_{2,\lsp ij}$ we have
\eqna{
    [a_{2,\lsp ij}] & =\tfrac{1}{180}(R^{\mu\nu\rho\sigma}R_{\mu\nu\rho\sigma}
    -R^{\mu\nu}R_{\mu\nu}+\tfrac52\llsp R^2+6\lsp\nabla^2 R)\delta_{ij}
    -\tfrac16(R+\nabla^2)X_{ij}+\tfrac12\llsp X_{ik}X_{ij}\,,
	\\
    [\partial_{\mu} a_{2,\lsp ij}] & =\tfrac{1}{180}(R^{\nu\rho\sigma\tau}\lsp
    \nabla_\mu R_{\nu\rho\sigma\tau} -R^{\nu\rho}\lsp\nabla_\mu
    R_{\nu\rho}+\tfrac52\llsp R\lsp\partial_\mu R
    +3\lsp\partial_\mu\nabla^2 R)\delta_{ij}\\
    &\quad-\tfrac{1}{12}\llsp\partial_\mu(R\llsp X_{ij})
    -\tfrac{1}{12}\llsp\partial_\mu\nabla^2 X_{ij}
    +\tfrac13(\tfrac12\llsp X_{ik}\lsp\partial_\mu X_{kj}
    +\partial_\mu X_{ik}\lsp X_{kj})\,,
    \\
    [\nabla^2a_{2,\lsp ij}]&=-(\tfrac{1}{540} \llsp K_{2} -\tfrac{1}{540}
    \llsp K_{3} -\tfrac{1}{1890}\llsp K_4-\tfrac{1}{630} \llsp K_{5}
    +\tfrac{1}{315} \llsp K_{6} -\tfrac{11}{3780}\llsp K_7
    +\tfrac{1}{189} \llsp K_{8} -\tfrac{1}{1008}\llsp K_9
    +\tfrac{1}{210} \llsp K_{10}\\
   &\quad\quad -\tfrac{1}{140} \llsp K_{11} -\tfrac{1}{140} \llsp K_{12}
   +\tfrac{1}{840} \llsp K_{13} +\tfrac{1}{420}\llsp K_{14}
   -\tfrac{3}{560} \llsp K_{15} -\tfrac{17}{3360}\llsp K_{16}
   -\tfrac{3}{280} \llsp K_{17} )\delta_{ij}\\
   &\quad-\tfrac{1}{90}(R^{\mu\nu\rho\sigma}R_{\mu\nu\rho\sigma}
   -R^{\mu\nu}R_{\mu\nu}+6\lsp\nabla^2 R
   +9\lsp\partial^\mu\hspace{-1pt} R\,\partial_\mu
   +3\lsp R^{\mu\nu}\lsp\nabla_\mu\partial_\nu
   +5\lsp R\lsp\nabla^2+\tfrac92\llsp(\nabla^2)^2)X_{ij}\\
   &\quad+\tfrac14(\tfrac13\llsp X_{ik}\lsp\nabla^2 X_{kj}
   +\nabla^2 X_{ik}\,X_{kj}+\partial^\mu X_{ik}\lsp\partial_\mu X_{kj})\,.
}[aIILimits]
Finally for $a_{3,\lsp ij}$ we have
\eqna{
	\left[ a_{3,\lsp ij} \right] & = \tfrac{1}{7!}
	\big(\tfrac{35}{9}\llsp K_1  -\tfrac{14}{3} \llsp K_{2}
    +\tfrac{14}{3} \llsp K_{3} +\tfrac{8}{9} \llsp K_{4}
    +\tfrac{8}{3} \llsp K_{5} -\tfrac{16}{3} \llsp K_{6}
    +\tfrac{44}{9} \llsp K_{7} -\tfrac{80}{9} \llsp K_{8} +11\llsp K_{9}\\
   &\quad\qquad-8 \llsp K_{10} +12 \llsp K_{11} +12 \llsp K_{12}
   -2 \llsp K_{13} -4 \llsp K_{14} +9\llsp K_{15}
   +\tfrac{17}{2} \llsp K_{16} +18 \llsp K_{17}\big)\delta_{ij}
	\\
	&\quad
	-\tfrac{1}{30}\big(
	\tfrac{1}{6}(R^{\mu\nu\rho\sigma}R_{\mu\nu\rho\sigma}
    - R^{\mu\nu} R_{\mu\nu}+ \tfrac{5}{2} R^2)
	\\
	& \quad\quad\quad\quad
    + \nabla^2 R + \partial^{\mu}\hspace{-1pt} R\,\partial_{\mu}
    + \tfrac{5}{6} R\lsp \nabla^2
	+ \tfrac{1}{3} R^{\mu\nu} \nabla_{\mu}\partial_{\nu}
	+ \tfrac{1}{2} (\nabla^2)^2\big) X_{ij}
	\\
	&\quad +
	\tfrac{1}{12} \big(
	R \lsp X_{ik} X_{kj}
	+ X_{ik} \lsp\nabla^2 X_{kj}
	+ \nabla^2 X_{ik}\lsp X_{kj}
	+ \partial^{\mu} X_{ik}\lsp \partial_{\mu} X_{kj}
	\big)
	\\
	&\quad - \tfrac{1}{6} X_{ik} X_{kl} X_{lj}
	\, .
}[aIIILimits]

It should be noted that all derivatives are evaluated inside the coincident
limits, i.e.\ the coincident limits of derivatives on the quantities are to
be taken, not the derivatives on the coincident limits of said quantities.
Actually, it is not difficult to convert between the two---the relevant
equation was given by Christensen~\cite{Christensen:1976vb} and was, in
fact, necessary for the computation of~(\ref{eq:twoLoopCT}).

\newsec{Coincident limits and divergences of products of
propagators}[appProp]

In this appendix we give the $\epsilon$ poles of the products of the
propagator pieces $G_n (x,x')$ and $R_n (x,x')$ found in \propExpansion. As
noted there, the full propagator in \greensfunction is regular for any $d$
at separate spacetime points. Explicitly,
\eqna{
	G_0 (x,x') & = \frac{\Gamma \left( \tfrac{1}{2}d - 1 \right) }{4\pi^{d/2}}
	\frac{\dVM}{(2\sigma)^{d/2-1}} \,,
\\
	G_1 (x,x') & =	\frac{\Gamma \left( \tfrac{1}{2}d - 2 \right) }{16\pi^{d/2}}
	\frac{\dVM}{(2\sigma)^{d/2-2}} \,,
\\
	R_n (x,x') & = \frac{\dVM}{4^{n+1}} \left( \frac{1}{\pi^{d/2}} \Gamma
	\left( \tfrac{1}{2} d - 1 - n \right) (2\sigma)^{n + 1 - d/2} -
	\frac{2}{\epsilon} \frac{(-1)^{n-1} \mu^{-\epsilon}}{\pi^3 (n - 2)!} (2\sigma)^{n - 2} \right)
	\, ,
}[singleProps]
for $n = 2, 3$.

However, upon taking products with other Green's functions, as in
\twoLoopGraph, short-distance singularities will arise that will have poles
in $\epsilon$, in accordance with \singleProps.  The associated relations
centrally depend on the coincident limit of inverse powers of $\sigma$ in
non-integer dimensions,
\eqn{
\frac{1}{\big(2\lsp\sigma(x,x')\big)^{\frac12(d-\delta)}}
\sim\frac{\mu^{-\delta}}{\delta}
\frac{2\lsp\pi^{d/2}}{\Gamma \left( \tfrac{1}{2} d \right)}
\delta^{\llsp d} (x,x')\,,}[sigmaPowers]
which is valid up to finite contributions as $x' \to x$. Here
$\delta\propto\epsilon$ and the $\mu^{-\delta}$ factor is inserted to
preserve dimensions. This dependence can be seen from the $\sigma$
dependence in equations in \singleProps. Varying powers of $\sigma$ will
arise depending on the product of propagators taken. A useful recursion
relation can be obtained by differentiation and the use of \fundamentalEqnA
and \fundamentalEqnB. It reads
\eqn{
	\left( \nabla^2 - Y  \right) \frac{\dVM}{\sigma^p} = p\lsp(2\lsp p+2-d)
	\frac{\dVM}{\sigma^{p+1}}\,.
}[Yrecursion]
Equation \Yrecursion can be used to obtain the poles in products
of propagators.  For example, if we multiply \sigmaPowers with $\dVM$, act
with $\nabla^2-Y$ and use \Yrecursion, we obtain
\eqn{\frac{\dVM}{(2\lsp\sigma)^{\frac12(d-\delta)+1}}\sim
\frac{\mu^{-\delta}}{\delta}\frac{\pi^{d/2}}{d\lsp\Gamma(\frac12 d)}
(\nabla^2-Y)\delta^{\llsp d}\,,}[DelsqY]
which is necessary for determining the $\epsilon$ poles in
$(G_0G_1G_1)(x,x')$ in $d=6-\epsilon$.

Using these methods we can now list the relevant products as
in~\cite{Jack:1985wf}:
\eqna{
	(G_0 G_0) (x,x') & \sim \frac{\mu^{-\epsilon}}{64\pi^3}
	\frac{1}{\epsilon} \frac{1}{3} \nabla^2 \delta^{\llsp d} (x,x') \,,
\\
	(G_0 G_1) (x,x') & \sim \frac{\mu^{-\epsilon}}{64\pi^3}
	\frac{1}{\epsilon} \delta^{\llsp d} (x,x') \,,
\\
	(G_0 G_1 G_1) (x,x') & \sim \frac{\mu^{-2\epsilon}}{(64\pi^3)^2}
	\frac{1}{\epsilon} \frac{1}{6} \left( \nabla^2 + \tfrac{1}{6} R \right)
	\delta^{\llsp d} (x,x') \,,
\\
	(G_1 G_1 G_1) (x,x') & \sim \frac{\mu^{-2\epsilon}}{(64\pi^3)^2}
	\frac{1}{\epsilon} \frac{1}{2} \delta^{\llsp d} (x,x') \,,
\\
	(G_0 G_1 R_2) (x,x') & \sim \frac{\mu^{-2\epsilon}}{(64\pi^3)^2} \left(
	\frac{1}{\epsilon^2} + \frac{1}{4}\frac{1}{\epsilon} \right)
	\delta^{\llsp d} (x,x') \,,
\\
	(G_0 G_0 R_3) (x,x') & \sim - \frac{\mu^{-2\epsilon}}{(64\pi^3)^2} \left(
	\frac{1}{\epsilon^2} + \frac{1}{4}\frac{1}{\epsilon} \right)
	\delta^{\llsp d} (x,x') \,,
\\
	(G_0 G_0 R_2) (x,x') & \sim \frac{\mu^{-2\epsilon}}{(64\pi^3)^2} \left(
	\frac{1}{\epsilon^2} - \frac{1}{12}\frac{1}{\epsilon} \right)
	\frac{1}{3} \left( \nabla^2 + \tfrac{1}{6} R \right)
	\delta^{\llsp d} (x,x')
	\, ,
}[productsProps]
where $\sim$ indicates that only the $\epsilon$ poles are considered on the
right side. In these expressions powers of $\dVM$ that appear in the
propagator products are commuted through to the delta function and then we
use $\dVM(x,x')\lsp\delta^{\lsp d}(x,x')=\delta^{\lsp d}(x,x')$.

For our purposes we also need to know the divergent behavior of the
products $(G_0 G_0 G_1) (x,x')$ and $(G_0 G_0 G_0) (x,x')$. Their
computation is performed after taking advantage of the fact that they
appear under $x,x'$-integrals, and thus we can integrate by parts at will.
For example, for $(G_0 G_0 G_1) (x,x')$ we need to find the poles in
$\big(\dVM\big)^3_{\phantom{3}}/(2\lsp\sigma)^{5-3\lsp\epsilon/2}$. From
\propExpansion and \DelsqY, \Yrecursion we see that we need
\eqn{\Delta_{\text{VM}}(\nabla^2-Y)^2\delta^{\lsp d}
=\Delta_{\text{VM}}\big[(\nabla^2)^2\delta^{\lsp d}
-\nabla^2(Y\delta^{\lsp d})-Y\nabla^2\delta^{\lsp d}
+Y^2\delta^{\lsp d}\big]\,.}[derDelta]
For the contribution of $(G_0 G_0 G_1) (x,x')$ to \twoLoopGraph we also
have $g_{ikl}(x)g_{jkl}(x')\lsp a_{1,\lsp ij}(x,x')$ under the
$x,x'$-integrals.  Following the procedure that led to \productsProps we
would now commute $\Delta_{\text{VM}}$ through the derivatives to the delta
function. This is rather tedious, so we choose here to integrate the
derivatives in \derDelta by parts instead. This way we are be able to do
the $x'$-integral which will force the coincident limits of the various
contributions that arise. The necessary results are then found in
\dVMLimits, \YLimits and \aILimits.

\end{appendices}

\bibliography{SixD}

\end{document}